\documentclass[10pt,namedreferneces]{SolarPhysics}
\usepackage[optionalrh,solaenum]{spr-sola-addons} 

\usepackage{graphicx}
\usepackage{color}
\usepackage{bm}

\newcommand{\BE}{\begin{equation}}
\newcommand{\EE}{\end{equation}}
\newcommand{\BA}{\begin{eqnarray}}
\newcommand{\EA}{\end{eqnarray}}

\newcommand{\BIT}{\begin{itemize}}

\newcommand{\EIT}{\end{itemize}}

\def \half {\textstyle{\frac{1}{2}}}

\begin{document}
\begin{article}
\begin{opening}
\title{Flux-Rope Twist in Eruptive Flares and CMEs: due to Zipper and Main-Phase Reconnection}

\author{E. R.~\surname{Priest}$^1$\sep 
D.W.~\surname{Longcope}$^2$}

\runningtitle{Flux-Rope Twist due to Zipper and Main-Phase Reconnection}
\runningauthor{E.R. Priest \&  D.W. Longcope}

\institute{$^1$ School of Mathematics and Statistics, University of St. Andrews, Fife KY16 9SS, Scotland, UK\\
$^2$ Dept. of Physics, Montana State University, Bozeman, MT, USA}

\begin{abstract} The nature of three-dimensional reconnection when a twisted flux tube erupts during an eruptive flare or coronal mass ejection is considered. The reconnection has two phases:  first of all, 3D ``zipper reconnection" propagates along the initial coronal arcade, parallel to the polarity inversion line (PIL); then subsequent  quasi-2D ``main phase reconnection" in the low corona around a flux rope during its eruption produces coronal loops and chromospheric ribbons that propagate away from the PIL in a direction normal to it.

One scenario starts with a sheared arcade: the zipper reconnection creates a twisted flux rope of roughly one turn ($2\pi$ radians of twist), and then main phase reconnection builds up the bulk of the erupting flux rope with a relatively uniform twist of a few turns.  A second scenario starts with a pre-existing flux rope under the arcade. 
Here the zipper phase can create a core with many turns that depend on the ratio of the magnetic fluxes in the newly formed flare ribbons and the new flux rope.
Main phase reconnection then adds a layer of roughly uniform twist to the twisted central core.  Both phases and scenarios are modeled in a simple way that 
assumes the initial magnetic flux is fragmented along the PIL. The model
uses conservation of magnetic helicity and flux, together with equipartition of magnetic helicity, to deduce the twist of the erupting flux rope in terms the geometry of the initial configuration.

Interplanetary observations show some flux ropes have a fairly uniform twist, which could be produced 
when the zipper phase and any pre-existing flux rope possess small or 
moderate twist (up to one or two turns). 
Other interplanetary flux ropes have highly twisted cores (up to five turns), which could be produced 
when there is a pre-existing flux rope and an active zipper phase that creates substantial extra twist.

\end{abstract}

\keywords{Sun: flares -- Sun: magnetic topology -- magnetic reconnection -- helicity}

\end{opening}

\section{Introduction} 
\label{sect_1}
The generally accepted overall scenario for an eruptive solar flare or coronal mass ejection that we adopt here may be  described as follows (\textit{e.g.}, \opencite{priest00}, \opencite{priest14a}).  During the pre-eruption phase, the magnetic configuration surrounding a solar prominence consists of a highly sheared magnetic structure, which gradually evolves through a series of force-free equilibria until it loses equilibrium or goes unstable and erupts.  The eruption starts slowly and then suddenly increases in speed.  Reconnection is driven below the erupting prominence, and the onset of reconnection is probably what allows the sudden increase in speed, since it cuts loose some of the overlying field lines that had previously been holding the prominence down.  (Other points of view include the suggestions that: eruptions are driven by Lorentz forces in the photosphere \cite{manchester04}; reconnection triggers the eruption \cite{shibata11}; and that reconnection is not necessary \cite{chen11}.)

The magnetic field in and around the prominence may consist of a highly sheared field or, more often, it is in the form of a  flux rope (a twisted magnetic flux tube).  The reconnection has the effect of either creating a new flux rope or  enhancing the flux and twist of a pre-existing flux rope \cite{gibson04,gibson06b}, since this is a natural consequence of three-dimensional reconnection \cite{priest16} and the erupting prominence is often observed to be much more highly twisted than was evident in its pre-eruptive state (\textit{e.g.},\ \opencite{mackay10a}; \opencite{mackay12b}).

Previously, there have been many numerical magnetohydrodynamic (MHD) studies of the build-up to, and initiation and evolution of,  flares and CMEs (see \textit{e.g.}, \opencite{priest14a}).  Ways of forming a flux rope include flux emergence \cite{archontis08a},  flux cancellation \cite{vanballegooijen89}, quasi-separator or slip-running reconnection \cite{aulanier10} and separator reconnection \cite{longcope07a}. Possible causes of the eruption include magnetic nonequilibrium \cite{priest90a,forbes91b,lin00}, kink instability \cite{hood79a,fan03,fan04,gibson04,torok04a}, torus instability \cite{kliem06,torok05} or breakout \cite{antiochos99a,devore08}.  In particular, physical mechanisms for the initiation and evolution of an eruption have been modelled numerically by \inlinecite{linker03}, \inlinecite{gibson08}, \inlinecite{fan10a}, \inlinecite{karpen12}, \inlinecite{aulanier12} and \inlinecite{titov12}. Also, the heating rate in flare loops produced by reconnection has been predicted \cite{longcope10a} and compared with {\it Solar Dynamics Observatory} observations \cite{qiu12a,li14}.

During the rise phase of a flare, H$\alpha$ ribbons form, and during the main phase the ribbons move apart and are joined by a series of hot flare loops, whose location rises as the flare progresses and whose shear becomes smaller \cite{moore01,qiu10,cheng12}.  As a coronal mass ejection (CME) propagates away from the Sun, it may be observed as an interplanetary coronal mass ejection (ICME) or magnetic cloud (MC) \cite{burlaga82a,burlaga95,lepping90,lepping97}, whose structure is that of a magnetic flux rope \cite{webb00,demoulin08,vourlidas14}. Magnetic clouds possess a low plasma beta and a strong magnetic field that rotates in direction. Indeed, it is likely that all ICMEs and MCs consist of flux ropes \cite{gopalswamy13}.  Initially, such interplanetary flux ropes were modelled as one-dimensional linear force-free fields or uniform-twist fields \cite{farrugia99,dasso06},  but more recently the observations have instead been compared with two-and-a-half dimensional Grad-Shafranov models \cite{hu02,hu14}.  The lengths of such flux tubes have been estimated from measurements of electron travel-times  \cite{larson97,kahler11} to be usually between 1 and 2 AU \cite{hu15}.

Several authors have compared the properties of interplanetary flux ropes with those of the erupting flux ropes at the Sun. \inlinecite{qiu07} showed that the poloidal flux ($F_{\rm p}$) in magnetic clouds is roughly equal to the total reconnected flux ($F_{\rm R}$) at the Sun ($F_{\rm p}\sim 1.1 F_{\rm R}^{0.8}$), which is strongly suggestive that interplanetary flux ropes are formed mainly by reconnection at the Sun during the initiation of the CME.  Also, the axial (toroidal) flux ($F_{\rm t}$) is less than the reconnection flux ($F_{\rm t}\sim 0.3 F_{\rm R}^{1.2}$).
During a CME, regions of coronal images darken, a process called ``coronal dimming" \cite{harrison00}, and such regions are thought to map the feet of the erupting flux rope and the surrounding region, since the erupting/opening process would allow plasma to escape outwards. \inlinecite{qiu07} found that the dimming flux ($F_{\rm d}$) is roughly equal to the toroidal flux in the MC ($F_{\rm d}\sim F_{\rm t}$) (see also \opencite{webb00}).

\inlinecite{hu14} extended the analysis to more events and
suggested a division into two possible types, which needs to be confirmed in future, bearing in mind the many different models that have been proposed (see, \textit{e.g.},  \opencite{dasso07} for references).
\inlinecite{hu14} found in their study that the twist either decreases with distance from the axis or is fairly small and constant.  Half of the interplanetary flux ropes (MCs) have a twist that is roughly constant  and is small (1.5--3 turns per AU), whereas the other half have a higher twist (up to about 5 turns per AU) that is concentrated in the core of the flux rope.  When the eruption is associated with a prominence (or filament), the mean twist tends to be lower. They also found that the sign of magnetic helicity in MCs is consistent with that of the flaring coronal arcade and confirmed that the poloidal flux in MCs is roughly equal to the measured reconnected flux in flares. 

One puzzle that we aim to consider in this paper is the cause of these variations in flux-rope twist, in particular what mechanisms could produce high twist in the core of some events but a low uniform twist in others. 
Another puzzle is an observational one that has been highlighted by \inlinecite{fletcher04} and \inlinecite{qiu09}, namely,  that the reconnection during a flare or CME often has two distinct phases with different characteristics 
\cite{yang09,qiu09,qiu10}, although sometimes the two phases overlap. During the first phase (the rise phase of an event), the flare brightening in H$\alpha$, EUV and hard X-rays starts at one point on each flare ribbon and the two points spread rapidly in the same direction  along the polarity inversion line (PIL) by what we call here ``zipper reconnection". By contrast, during the main phase, the flare ribbons move outwards in a direction normal to the PIL, by what we call ``main-phase reconnection". The speed of spread along the PIL lies between 3 and 200 km s$^{-1}$, whereas the speed normal to the PIL starts very fast (at sometimes 100 km s$^{-1}$) and slows down later to as small as 1 km s$^{-1}$.   Sometimes, rather than being unidirectional, the flare kernel motions are bidirectional in the sense that they may spread out from one point in both directions along the ribbons \cite{su07,yang09}. Sometimes shear motion is observed, in the sense that conjugate footpoints move in opposite directions: at the same time,  the angle between the footpoints and the polarity inversion line starts out small and increases as the flare progresses and the corresponding flare loops rotate. This, however, is different from the zipper effect of footpoints moving in the same direction that we are modelling here. Shear motion is usually thought to be associated with the progression of reconnection from low highly sheared fields to high less-sheared fields that we are describing here as main-phase reconnection.

Various studies have been made of the spread of reconnection along the PIL \cite{vorpahl76,kitahara90,fletcher03,krucker03,fletcher04,bogachev05,tripathi06,su07,li09,qiu10,cheng12}. For example, \inlinecite{li09} found half of the events in their sample  have a speed of 10--40 km s$^{-1}$ in one direction (opposite to \textbf{\textit {j}}$_{\parallel}$) along the PIL, whereas the other half have speeds of 100--200  km s$^{-1}$ that are either bidirectional (in both directions from the starting point) or are in the same direction as \textbf{\textit {j}}$_{\parallel}$.
\inlinecite{aulanier00a} describe the 1998 Bastille Day flare, and suggest the standard 2D model that the formation of the flare ribbons is due to relaxation of field lines that have been blown open by an eruption. In our scenario, this only refers to the transverse motion of the ribbons, whereas the  formation is associated with the  zipping of a new flux rope (usually around a pre-existing one).

Several 3D computations of flux rope eruptions followed by reconnection have been described, but none of them, to our knowledge, have focused on the difference in the nature of the reconnection and the motion of chromospheric brightenings during the rise and main phases of the flare. For example, \inlinecite{amari03a} model the formation and eruption of a twisted flux rope.  They show that the magnetic helicity remains constant and describe reconnection in the main phase. \inlinecite{fan03}, \inlinecite{fan04}, \inlinecite{fan07a} and \inlinecite{fan10a} show how eruption occurs by the kink or torus instability and how the reconnection builds up twist in the flux rope during the main phase.

Furthermore, \inlinecite{devore08} model an active region containing a coronal null and show how multiple eruptions can be driven by the magnetic breakout scenario. Also, \inlinecite{masson09a}, \inlinecite{masson12a} and  \inlinecite{aulanier10} have proposed that the flipping (or slipping) motion, which is a natural feature of 3D reconnection at a null point, separator or quasi-separator, can account for motion of brightenings along ribbons.  This seems an excellent explanation for the shear motions in opposite directions reported in some flares by \inlinecite{su07} and \inlinecite{yang09}, but this is different from the case of zipper motions in the same direction that we are modelling here.

New features we bring here are an explanation for the initial motion of flare brightenings parallel to the PIL and a suggestion as to how the structure of the resulting twist in the erupting flux rope arises and is related to the initial configuration and the two reconnection phases.

Three explanations have previously been offered to explain the spreading of flare emission along the polarity inversion line.  
\inlinecite{bogachev05} and \inlinecite{liu09} described observations of hard X-ray sources and suggest that the flare acceleration region is moving along the PIL (with which we agree), but they give no description of the nature or consequences of the reconnection. \inlinecite{desjardins09} modelled the topology of a flare with null points in which the hard X-ray sources move along a series of spines by separator reconnection. Finally, \inlinecite{li14b}, \inlinecite{li15}, \inlinecite{dudik14} and \inlinecite{dudik16} considered flares without null points, but modelled the quasi-topology and the reconnection in terms of quasi-separator (or slipping) reconnection.  Our ideas build on  the two latter ideas by using concepts of magnetic helicity conservation to deduce the effect of either separator or quasi-separator reconnection on the buildup of twist in the erupting flux rope. We deal with separator reconnection, but, if the null points are replaced by weak-field regions, it simply becomes quasi-separator reconnection. In both cases, the reconnecting field lines experience flipping motions.

We find that our model shows how zipper reconnection acting on a sheared arcade followed by main-phase reconnection produces a core twist of at most 2$\pi$ inside a region of uniform twist, and this can naturally explain observed cases of low uniform twist.  By contrast, zipper and main phase reconnection acting on a  twisted pre-existing flux rope can naturally give rise to erupting flux ropes with high core twist. 

Section \ref{sect_2} describes the initial state of the model, and Section \ref{sect_3} uses concepts of flux and helicity conservation and equipartition proposed in \inlinecite{priest16} to model elementary reconnection events of two types, namely a ``simple zippette" between a pair of flux tubes in a sheared arcade and a ``helical zippette" when the arcade overlies an initial flux rope. Such helicity-conserving reconnection is also important in tokamaks and for heating  multi-threaded coronal loops \cite{browning16a,hood16a}. Section \ref{sect_4}  presents the model for 3D zipper reconnection, 
both for an initial sheared arcade and for an initial arcade containing a flux rope,
while Section \ref{sect_5} discusses the subsequent  main-phase reconnection.

\section{Setting Up the Model -- the Initial State} 
\label{sect_2}

Our approach is to set up a simple model to predict the twist in an erupting flux rope in terms of the properties of the pre-eruptive state by imposing conservation of magnetic flux and magnetic helicity and equipartition of magnetic helicity.  In future, it is hoped this can be applied both to observations and to computational models. The extra constraint of energy was suggested by \inlinecite{linton01} and considered briefly by \inlinecite{priest16}, but needs computational modelling and so is outside the scope of the present paper.
When energy effects are included they may allow the thermodynamic properties to be determined and they also have the potential to provide extra constraints and so rule out some scenarios that are energetically unfavourable.
\begin{figure}[h]
{\centering
 \includegraphics[width=12cm]{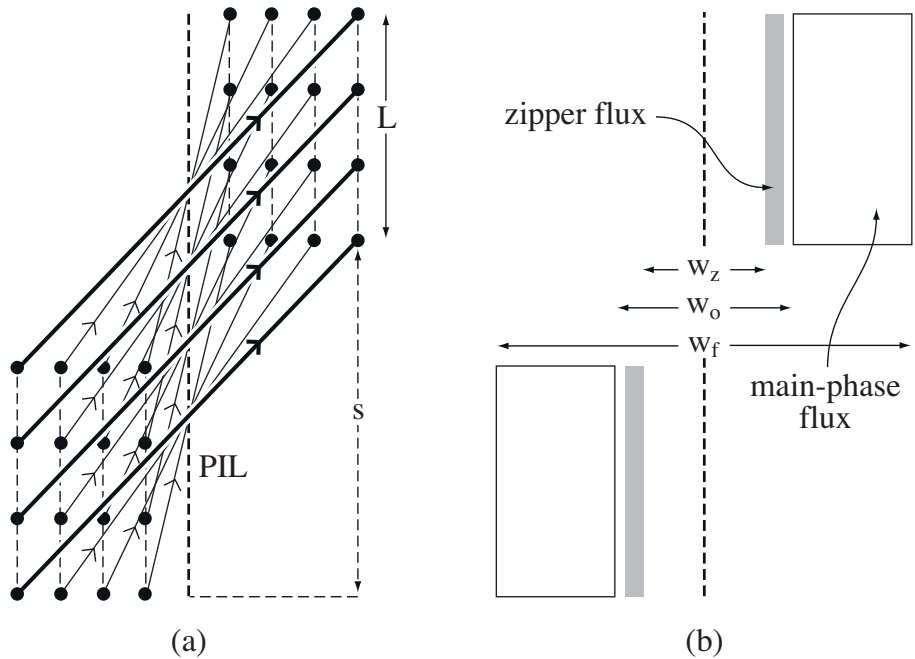}
\caption{(a) The magnetic field lines of a sheared arcade viewed from above, whose footpoints are indicated by dots and are located either side of a polarity inversion line (PIL). (b) The flux regions viewed from above taking part in the zipper and main phases of reconnection.}
\label{fig1}}
\end{figure}

We consider first a pre-eruptive sheared coronal arcade, the shear of whose field lines decreases as one moves away from the polarity inversion line (PIL) (Figure \ref{fig1}a).  The length of the arcade is $L$ and its width is $w_f$, while the shear of the arcade as a whole is $s$.  
We assume, following the standard procedure in the well-established ``magnetic charge topology" approach \cite{sweet58a,molodensky77,welsch99,brown99a,longcope02,beveridge02,longcope05b}, that the magnetic field sources in the photosphere are modelled as a series of discrete sources rather than a continuous distribution. As an example, Figure \ref{fig1}a shows a set of four rows of four sources.  If the discrete sources become continuous or are placed below the photosphere rather than on it, many of the null points and separators become weak-field regions and quasi-separators, but the nature of the reconnection is very similar \cite{seehafer86,gorbachev88c,demoulin92a,priest95a}.

We suppose that three-dimensional reconnection takes place in two phases. During the rise phase, reconnection  starts in general at any point within the innermost field lines and spreads in both directions along the arcade parallel to the PIL by  zipper reconnection (Section \ref{sect_4}). For simplicity, we start by modelling the case where it starts at one end and proceeds in one direction. The flux that takes part in the zipper reconnection occupies two narrow regions either side of the PIL separated by a distance $w_z$ (Figure \ref{fig1}b).  

During the main phase of the flare, reconnection takes place by  main-phase reconnection (Section \ref{sect_5}), in which the reconnection spreads out laterally in a direction perpendicular to the PIL. The magnetic flux that takes part in this phase occupies two strips whose inner and outermost parts are separated by $w_0$ and $w_f$, respectively (Figure \ref{fig1}b).  

As a second case, we consider instead as the initial state in Section  \ref{sect_4.2} a sheared coronal arcade overlying a flux rope, since this is thought in many cases to be more representative of the pre-eruptive configuration around a prominence that erupts to give the flare or coronal mass ejection.

\section{Elementary Reconnection Events} 
\label{sect_3}

\subsection{\bf A Simple Zippette: a Single Reconnection Event of a Sheared Arcade} 
\label{sect_3.1}
\begin{figure}[h]
{\centering
 \includegraphics[width=12cm]{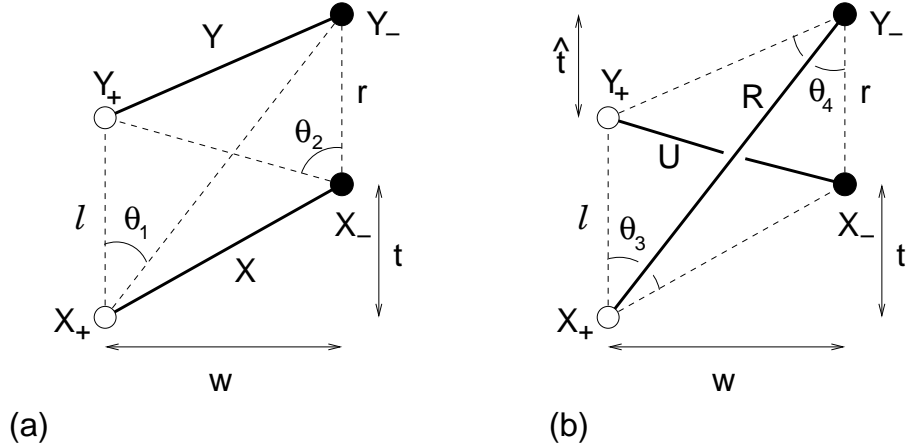}
\caption{The geometry for quantifying a ``simple zippette", a typical single reconnection event in the zipper phase of a sheared coronal arcade.  (a) The initial configuration ($XY$), in which sources X$_+$ and X$_-$ are connected by one flux tube ($X$), and  Y$_+$ and Y$_-$ are joined by a second tube ($Y$).  (b) The final configuration ($UR$), in which the initial flux tubes have reconnected to form a flux rope ($R$) connecting X$_+$ to Y$_-$.  This lies above a second underlying flux loop ($U$), which connects Y$_+$ to X$_-$.  Interior angles ($\theta_j$) are indicated for both configurations.}
\label{fig2}}
\end{figure}
We assume that the zipper reconnection phase of a sheared arcade is composed of multiple individual reconnection events, called ``simple zippettes", between tube pairs of equal flux.  Figure \ref{fig2} depicts a single generic event viewed from above in which flux tubes $X$ and $Y$ reconnect to form a new pair of flux tubes, $U$ and $R$.  All sources, and therefore all flux tubes, have exactly the  same magnetic flux, $F$.  The positive sources (X$_+$ and Y$_+$) are separated by some distance $\ell$, while the negatives (X$_-$ and Y$_-$) are separated by a possibly different distance ($r$) along a parallel line -- shown here as vertical.  The two opposing lines are themselves separated by a perpendicular distance $w$, and X$_-$ is located a distance $t$ further along its line than X$_+$.  The figure shows a case with $t>0$, but the generic calculation applies to $t<0$ as well.
Figure \ref{fig2}a gives the general directions of the flux tubes linking the sources initially, but, as we shall discuss in the next subsection they are made up of field lines that are curved and not parallel so that they can reconnect.  Furthermore, we are assuming here in Figure \ref{fig2}b that $R$ lies above $U$, since we are considering a scenario in which long erupting flux ropes are created. An argument to suggest that this is energetically allowable was presented briefly in \inlinecite{priest16}.

The process of reconnection converts the initial configuration, designated $XY$, to a final configuration ($UR$) with a flux rope $R$ above an underlying arcade loop $U$ (see Figure\ \ref{fig2}).  The reconnection process is assumed to conserve helicity, but changes the partitioning between self-helicity and mutual helicity.  The self-helicity of a single flux tube twisted by total angle $\Phi$ (positive for right-handed twist) is $H^{\rm s} =F^2\Phi/2\pi$.

The mutual helicity of a particular configuration is proportional to the signed sum of its two interior angles, designated 
$\theta_j$ in Figure\ \ref{fig2} \cite{demoulin06b}.  One interior angle is computed for each footpoint of the overlying flux tube.  When the flux tubes are separated and neither is above the other, then either may be selected as the first flux tube -- tube $X$ has been chosen in Figure\ \ref{fig2}a.  The interior angle for a given footpoint, the angle's vertex, is that subtended by the two footpoints of the other flux tube.  A positive sign is assigned if the footpoints of that tube appear counter-clockwise when proceeding from the source with the same sign as the vertex.  Thus, the interior angle of X$_+$, designated $\theta_1$ in Figure\ \ref{fig2}a, contributes to the mutual helicity with a negative sign, since going from Y$_+$ to Y$_-$ is a clockwise direction when viewed from X$_+$.  

All angles are assigned values in the range $0\le\theta_j\le\pi$ and the sign discussed above is used when computing the mutual helicity.  For the orientations  in Figure\ \ref{fig2} we find mutual helicities
\begin{equation}
  H^{\rm m}_{XY} ~=~ {F^2\over\pi}(\, \theta_2 - \theta_1)~,~~~ H^{\rm m}_{UR} ~=~ -{F^2\over\pi}(\, \theta_3 + \theta_4)~,
  	\label{eq:HmXY}
\end{equation}
in agreement with similar expressions in \inlinecite{priest16}, who treated the special case ($r=l$) of a parallelogram.  Having accounted for their signs in Equation (\ref{eq:HmXY}), all angles  can be found from the geometry of the figure.  The angles in the $XY$ configuration, shown in Figure \ref{fig2},  are
\begin{equation}
  \tan\theta_1 ~=~{w\over r+t} ~=~{1\over \bm\bar{r}+\bm\bar{t}} ~,~~\tan\theta_2 ~=~{w\over \ell-t} ~=~{1\over \bm\bar{\ell}-\bm\bar{t}} ~,
  	\label{eq:th12}
\end{equation}
where we  introduce dimensionless distances normalised with respect to the polarity separation $w$, namely, $\bm\bar{t}=t/w$, $\bm\bar{r}=r/w$, and $\bm\bar{\ell}=\ell/w$.  The angles in the $UR$ configuration are 
\begin{equation}
  \tan\theta_3 ~=~{1\over \bm\bar{t}} ~,~~\tan\theta_4 ~=~{1\over \bm\bar{t}+\bm\bar{r}-\bm\bar{\ell}} ~.
  	\label{eq:th34}
\end{equation}
The figures depict a case where $t>0$, so the overlying flux rope $R$, is longer than $U$, and both $\theta_3$ and $\theta_4$ are acute angles.  Inspection of the figures reveals, however, that  (\ref{eq:th12}) and (\ref{eq:th34}) are valid even when $t<0$, in which case at least $\theta_3$  becomes obtuse ($\tan\theta_3<0$).  Since the quadrilateral $Y_+X_+X_-Y_-$ retains its orientation for all values of $t$, none of the angles change their sense, and expression (\ref{eq:HmXY}) remains valid when $t<0$.

A reconnection event begins in configuration $XY$, with flux in tubes $X$ and $Y$ being twisted by, say, 
angles $\Phi_X$ and $\Phi_Y$, respectively.  The tubes $U$ and $R$ in the final configuration are also be twisted in general.  We  assume that the reconnection process contributes equal self-helicity to each, an assumption we call 
``helicity equipartition".  This means that $\Phi_U=\Phi_R$.  The conservation of total helicity therefore implies
\begin{equation}
  {F^2\over2\pi}(\Phi_X+\Phi_Y) ~+~ H^{\rm m}_{XY} ~=~ {F^2\over\pi}\Phi_R ~+~ H^{\rm m}_{UR} ~.
  \label{eqnconshel}
\end{equation}
Substituting the mutual helicity expressions from Equations (\ref{eq:HmXY}), we are able to deduce the final twist in terms of the properties of the initial configuration
\begin{equation}
  \Phi_R ~=~ \hbox{${1\over2}$}(\Phi_X+\Phi_Y) + (\theta_3+\theta_4+\theta_2-\theta_1) ~=~ \hbox{${1\over2}$}
  (\Phi_X+\Phi_Y) +\Delta\Phi^{\rm m} ~,
\end{equation}
where $\Delta\Phi^{\rm m}$ is the twist contribution due to a change of mutual helicity.

For configurations with the qualitative appearance of Figure\ \ref{fig2}, 
$\Delta\Phi^{\rm m}=\theta_3+\theta_4+\theta_2-\theta_1$ is positive.  This reflects the fact that configuration $UR$ has a single crossing in the negative sense \cite{berger93,berger06}, which makes its mutual helicity negative [see Equation \ (\ref{eq:HmXY})].  A positive self-helicity, \textit{i.e.},\  right-handed twist, is required to compensate and thereby conserve total helicity.  

In the special case when the quadrilateral $Y_+X_+X_-Y_-$ is a rectangle ($t=0$, $\ell=r$) it is evident that $\theta_3=\theta_4=\pi/2$ and $\theta_2=\theta_1$, so that reconnection adds exactly one half-twist to each tube: $\Delta\Phi^{\rm m}=\pi$. 
This startling fact is a natural and elegant consequence of conversion (with equipartition) of mutual helicity to self-helicity in the special geometry when the footpoints lie on the vertices of a rectangle.  In this case, the initial and final mutual helicities of the two tubes become simply from Equation (\ref{eq:HmXY})
\begin{equation}
  H^{\rm m}_{XY} ~=~ 0,~~~ H^{\rm m}_{UR} ~=~ -F^2,
\end{equation}
whereas the sum of the final self-helicities of the two tubes becomes $(F^2/\pi)\Phi_R$,
so that, when the initial self-helicities vanish ($\Phi_X=\Phi_Y=0$), Equation (\ref{eqnconshel}) implies $\Phi_R=\pi$. This result was confirmed numerically in the three-dimensional resistive MHD experiment of \inlinecite{linton03}.
When the sources do not form a rectangle, then $\Delta\Phi^{\rm m}\neq \pi$, but  we shall find that, even so, often it does not depart much from $\pi$: for example, when ${\bm\bar L}<6\sqrt 2$ the twist lies between $\pi/2$ and  $3\pi/2$ (see Appendix).

Explicit dependence on angles can be eliminated by using Equations\ (\ref{eq:th12}) and (\ref{eq:th34}) to give 
\begin{eqnarray}
\tan(\Delta\Phi^{\rm m}) &=& {\bm\bar{r}\bm\bar{\ell}(2\bm\bar{t}+\bm\bar{r}-\bm\bar{\ell})\over
  \bigl[(\bm\bar{r}+\bm\bar{t})(\bm\bar{\ell}-\bm\bar{t})+1\bigr]\,\bigl[\bm\bar{t}(\bm\bar{t}+\bm\bar{r}-\bm\bar{\ell})-1\bigr]-
  (2\bm\bar{t}+\bm\bar{r}-\bm\bar{\ell})^2} ~. \label{eq:dPhi} 
\end{eqnarray}
A more tractable expression can be obtained by introducing the variable $\bm\hat{t}=\bm\bar{t}+\bm\bar{r}-\bm\bar{\ell}$, whose geometric significance is indicated in Figure\ \ref{fig2}b.  Using this, Equation  (\ref{eq:dPhi}) can be written
\begin{eqnarray}
\tan(\Delta\Phi^{\rm m}) &=& -{\bm\bar{r}\bm\bar{\ell}(\bm\bar{t}+\bm\hat{t})\over
  (\bm\bar{t}\bm\hat{t} - 1-\bm\bar{r}\bm\bar{\ell})\,(\bm\bar{t}\bm\hat{t} - 1)
  +(\bm\bar{t}+\bm\hat{t})^2} ~. \label{eq:dPhi2}
\end{eqnarray}
The appropriate branch of tan can be assigned using the fact, established above, that $\Delta\Phi^{\rm m}\to\pi$ as the configuration becomes rectangular so that  $\bm\bar{t}$ and $\bm\hat{t}$ tend to $0$.

\subsection{\bf The Nature of Simple Zippette Reconnection} 
\label{sect_3.2}
\begin{figure}[h]
{\centering
 \includegraphics[width=12cm]{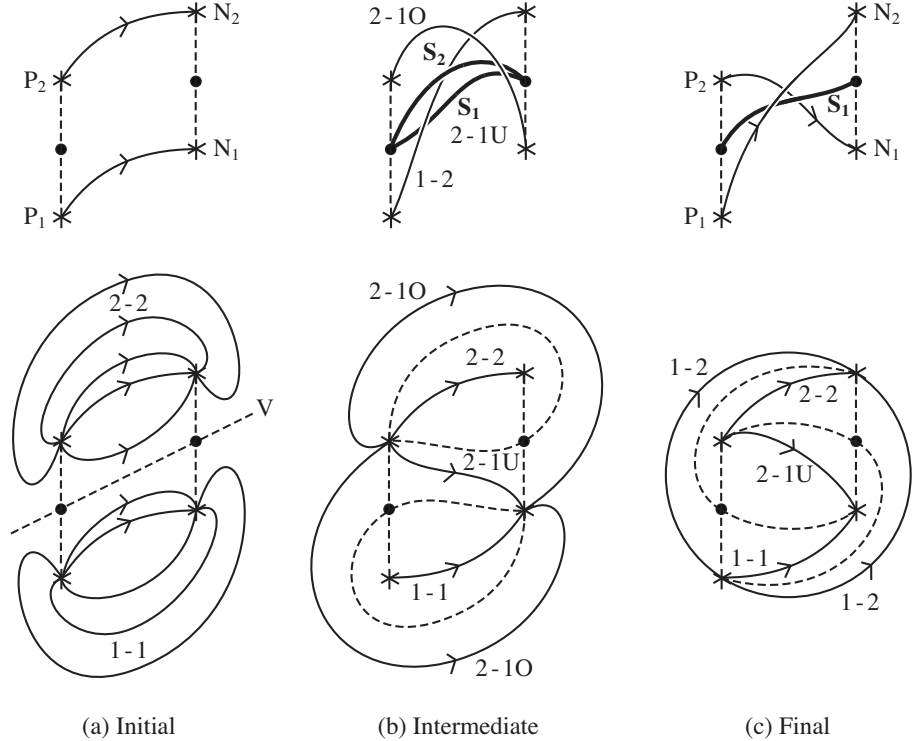}
\caption{The topology of double separator reconnection during a simple zippette with positive  (P$_1$ and P$_2$) and negative (N$_1$ and N$_2$) flux sources, passing from (a) the initial topology through (b) an intermediate phase to (c) the final topology. Null points lying midway between P$_1$ and P$_2$ and between N$_1$ and N$_2$ in the photospheric plane are indicated by large dots. The diagrams on the top line represent the overall topology, while those below show the detailed topology in the photospheric plane with dashed curves indicating the intersections of separatrix surfaces with the photosphere.  (a) Initially, there is flux (1-1) joining P$_1$ to N$_1$ and flux (2-2) joining P$_2$ to N$_2$, separated by a vertical separatrix surface V. (b) In the intermediate stage during reconnection, there are two separatrix curves (S$_1$ and S$_2$) joining the null points (see above). Also, some of the fluxes 1-1 and 2-2 have been converted into 
overlying flux (2-1O) joining P$_2$ to N$_1$ but lying over separatrix S$_2$,  underlying flux (2-1U) joining P$_2$ to N$_1$ but lying underneath separatrix S$_1$ and flux (1-2) that joins P$_1$ to N$_2$ and passes through the ring formed by separators S$_1$ and S$_2$. (c) In the final stage the reconnection has been completed and separator S$_2$ has disappeared to infinity so that no flux 2-1O remains}.
\label{fig3}}
\end{figure}
\begin{figure}[h]
{\centering
 \includegraphics[width=12cm]{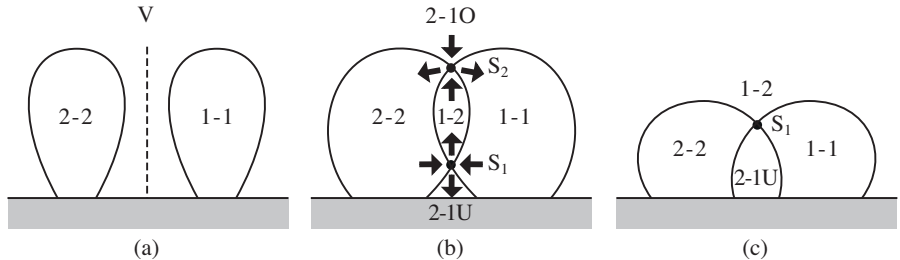}
\caption{A vertical section through the separator reconnection process during zippette reconnection that is described in Figure \ref{fig3}. (a) Two flux regions 1-1 and 2-2 filling the space either side of a vertical separatrix surface (V). (b) Separator reconnection at two separators S$_1$ and S$_2$ transfers flux between 1-1, 2-2 and new flux regions 1-2, 2-1U and 2-1O. (c) The final state possesses a separator S$_1$ and flux regions 1-2 and 2-1U in addition to the two initial regions}.
\label{fig4}}
\end{figure}

The way in which reconnection takes place in a simple zippette is by double separator reconnection. This process is more complex than may be at first thought, because of the nature of the topology and its changes, as has been discussed previously by, e.g., \inlinecite{brown99a}, \inlinecite{brown99b}, \inlinecite{parnell08a}, \inlinecite{parnell10a}, \inlinecite{haynes07a}, \inlinecite{longcope96a} and \inlinecite{longcope01}.  Suppose we have two positive flux sources (P$_1$ and P$_2$) and two negative sources (N$_1$ and N$_2$), and that initially all the flux from P$_1$ links to N$_1$, while all the flux from P$_2$ links to N$_2$. The resulting magnetic topology is sketched in Figures \ref{fig3}a and \ref{fig4}a, as viewed from above and from the right, respectively. The flux (1-1) from P$_1$ to N$_1$ is separated from the flux (2-2) from P$_2$ to N$_2$ by a vertical separatrix surface (V) which contains two null points (indicated by large dots in Figure \ref{fig3}). Thus, the whole region to the one side of V contains flux 1-1, while the whole region on the other side contains flux 2-2.

Suppose for simplicity that, in the final state when reconnection has been completed, all the flux (1-2) that goes from P$_1$ passes over the underlying flux (2-1U) from P$_2$ to N$_2$. (It is possible that such a state is not reached and that instead the final state has the form of one of the intermediate states, but we shall not discuss such a situation here.) However, the presence  of some flux (2-2) from P$_2$ to N$_2$ and also flux (1-1) from P$_1$ to N$_1$ implies that the footprint of the topology is as sketched in Figure \ref{fig3}c when viewed from above, with the intersections of the separatrix surfaces with the photosphere indicated by dashed curves. The separatrix surfaces are in the form of two domes which intersect in a separator curve (S$_1$) joining one null point to the other and lying above P$_2$N$_1$ but below P$_1$N$_2$. The feet of one of the domes pass through the sources P$_1$ and P$_2$ and both null points, while the feet of the other dome pass through N$_1$, N$_2$ and both nulls. In a vertical plane that passes through both domes (Figure \ref{fig4}c) the intersection with the separator is indicated by a large dot S$_1$.

As discovered in previous numerical experiments \cite{parnell08a} and analyses of topological bifurcations \cite{brown99a,brown99b}, it is clear that the path from the initial to the final state involves a series of complex intermediate states having the form shown in Figures \ref{fig3}b and \ref{fig4}b, in which the two separatrix domes intersect in two separators (S$_1$ and S$_2$) which link the two null points. The flux linking P$_2$ to N$_1$ consists of a underlying and an overlying part, one of which (2-1U) lies underneath separator S$_1$, while the other (2-1O) overlies separator S$_2$.  

The two separators S$_1$ and S$_2$ form by a global separator bifurcation \cite{brown99a} as the two flux surfaces touch and intersect one another. In the final state the upper separator has disappeared to infinity.  During separator reconnection at the lower separator S$_1$, flux is transferred from regions 1-1 and 2-2 into 2-1U and 1-2, while reconnection at the upper separator S$_2$ transfers flux from 2-1O and 1-2 into 1-1 and 2-2.

It should be noted that instead of regarding the flux sources  for simplicity as point sources,  we could regard them as finite sources or as continuous sources, so that the photospheric field is continuous.  In this case, some of the null points, separatrix surfaces and separators disappear, but remnants of them remain as weak-field regions, quasi-separatrix layers and quasi-separators (or hyperbolic flux tubes) \cite{priest95a,demoulin96a,demoulin97d,aulanier05a,aulanier06a,aulanier07}.

\subsection{\bf A Helical Zippette: a Reconnection Event in an Arcade with a Flux Rope} 
\label{sect_3.3}

\begin{figure}[h]
{\centering
 \includegraphics[width=12cm]{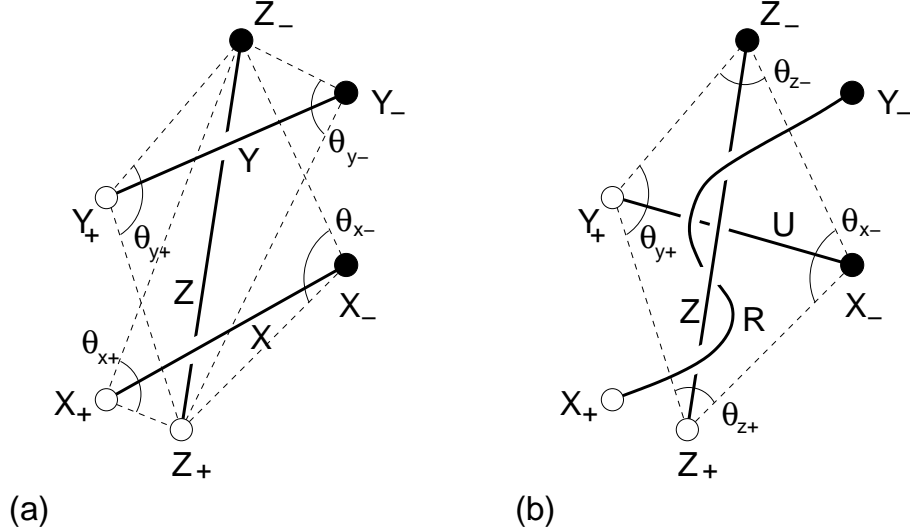}
\caption{The geometry for quantifying a ``helical zippette", a typical single reconnection event between flux tubes $X$ and $Y$ in the zipper phase of a coronal arcade that overlies an initial flux rope $Z$. The initial and final states are shown in (a) and (b), respectively.  The reconnection occurs beneath $Z$, and thus produces an underlying tube, $U$.  The other  tube that is produced by the reconnection ($R$) overlies both $U$ and $Z$, and wraps one entire time around $Z$.  The interior angles with flux rope $Z$ are indicated.}
\label{fig5}}
\end{figure}
We assume that the zipper reconnection phase of a coronal arcade that overlies an initial flux rope $Z$ is composed of multiple individual reconnection events, called ``helical zippettes", between tube pairs of equal flux. Figure \ref{fig5} depicts a single generic event viewed from above in which flux tubes $X$ and $Y$ reconnect to form a new pair of flux tubes, one of which ($U$) lies beneath rope $Z$ while the other ($R$) wraps around $Z$.

This event is topologically identical to the simple case shown in Figure\ \ref{fig2}, except for the presence of the third tube $Z$.  Since that tube does not actually participate in the reconnection process, it has no effect on the outcome. Thus, all results from Section \ref{sect_3.1} apply equally to the present case, irrespective of the presence of flux tube $Z$.  This fact is not, however, immediately obvious owing to the way $Z$ appears to impose itself into the configuration, such that flux tube $R$ ends up completely wrapped around it.  We thus demonstrate our conclusion by computing the mutual helicities before and after the event.  This also shows that tube $R$ must end up above tube $Z$ and must wrap about it once, as indicated in Figure\  \ref{fig5}.

With three flux tubes there are three distinct pairings for which mutual helicity contributions must be calculated.  When considering the pairs $XY$ (before) and $UR$ (after), it is clear that their  topology is identical with those in Figure\ \ref{fig2}: $X$ and $Y$ are separate, while $R$ overlies $U$.  The mutual helicities are therefore given by Equation  (\ref{eq:HmXY}), with interior angles given by Equations\ (\ref{eq:th12}) and (\ref{eq:th34}).  

It remains, then, to compute the mutual helicity contributions from the other pairings.  The interior angles for these pairings are shown in Figure\ \ref{fig5}, with natural labels $\theta_{x+}$, $\theta_{x-}$, etc.  In terms of these, the mutual helicities of the initial configuration are
\begin{equation}
  H^{\rm m}_{XZ} ~=~{FF_z\over \pi}\Bigl(\, \theta_{x+} ~+~ \theta_{x-}\,\Bigr) ~,~~
  H^{\rm m}_{YZ} ~=~{FF_z\over \pi}\Bigl(\, \theta_{y+} ~+~ \theta_{y-}\,\Bigr) ~,
  	\label{eq:HmXZ}
\end{equation}
where $F_z$ is the flux of rope $Z$, and the positive senses of all angles can be seen from Figure\ \ref{fig5}.

The final flux tube $R$ connects footpoints X$_+$ and Y$_-$ by going over rope $Z$.  The footpoints of this flux tube are used to compute the helicity, and they have 
the same interior angles ($\theta_{x+}$ and $\theta_{y-}$) as used for the initial state.  However, since the tube R wraps $Z$ one entire time, in the right-handed sense, as well as lying above it, the mutual helicity is
\begin{equation} 
  H^{\rm m}_{RZ} ~=~{FF_z\over \pi}\Bigl(\, \theta_{x+} ~+~ \theta_{y-}~+~2\pi \,\Bigr) ~.
  	\label{eq:HmRZ}
\end{equation}
Since flux tube $Z$ lies above $U$, we must use its foot-points ($Z_+$ and $Z_-$) for the interior angles to compute the mutual helicity of $UZ$
\begin{equation}
  H^{\rm m}_{UZ} ~=~ -{FF_z\over \pi}\Bigl(\, \theta_{z+} ~+~ \theta_{z-}\,\Bigr) ~,
\end{equation}
with a sign change due to the senses of the angles.  The interior angles of the quadrilateral $X_-Z_-Y_+Z_+$ must sum to $2\pi$, so it follows that $\theta_{z+}+\theta_{z-}=2\pi-\theta_{x-}-\theta_{y+}$, which implies that
\begin{equation}
  H^{\rm m}_{UZ} ~=~{FF_z\over \pi}\Bigl(\, \theta_{x-} ~+~ \theta_{y+}~-~2\pi \,\Bigr) ~.
\end{equation}
We therefore see that the mutual helicity contributions of $Z$, albeit non-trivial, do not change through the reconnection process
\[
  H^{\rm m}_{XZ}+H^{\rm m}_{YZ} ~=~H^{\rm m}_{RZ}+H^{\rm m}_{UZ} ~.
\]

Had we stipulated at the outset that flux tube $Z$ cannot affect the result, the foregoing logic would have led us to the realization that in order for one of the resulting flux tubes ($U$) to lie underneath $Z$, it is necessary for the other to not only overlie it, but also to wrap that tube once completely in the right-handed sense.  Following the same logic, we can generalize to the case where flux ropes $X$ and $Y$ initially wrap around $Z$ some numbers, $N_x$ and $N_y$, of times, respectively, in the right-hand sense.  This would add $2\pi N_x$ and $2\pi N_y$ to the angles in the mutual helicities of Equation  (\ref{eq:HmXZ}).  If flux rope $U$ does not wrap $Z$, then flux rope $R$ must wrap it a total of $N_x+N_y+1$ times in order to conserve mutual helicity.

The presence of flux rope $Z$ does not, therefore, affect the change in mutual helicity during the reconnection event.  Nor does it affect the change in self-helicity since its twist and its flux are preserved during the process.  The only changes are therefore those same changes accounted for in Section \ref{sect_3.1}.  As a result, flux ropes $R$ and $U$ have identical twists.  That twist is the mean, $(\Phi_X+\Phi_Y)/2$, plus a contribution due to the reconnection given by Equations\ (\ref{eq:dPhi}) or (\ref{eq:dPhi2}).  Thus, flux rope $R$ has internal twist in addition to being wrapped around $Z$.

The reconnection, however,  creates a composite structure (consisting of the sum of $R$ and $Z$) whose self-helicity combines self and mutual contributions of its components.  
Suppose we combine the new flux rope ($R$) and the initial one ($Z$) to give an erupting flux rope of flux 
\begin{equation}
F_{ER}=F+F_Z~.
\end{equation}
The twist ($\Phi_{ER}$) of this composite structure is determined by the fact that its self-helicity should be the sum of the self-helicities of $R$ and $Z$ together with their mutual helicity ($H^{\rm m}_{RZ}$) from Equation  (\ref{eq:HmRZ}), namely,
\begin{equation}
\frac{\Phi_{ER}F_{ER}^2}{2\pi}=\frac{\Phi_{R}F^2}{2\pi}+\frac{\Phi_{Z}F_{Z}^2}{2\pi}+\frac{FF_Z}{\pi}(\theta_{x+} + \theta_{y-}+2\pi),
\end{equation}
so that 
\begin{equation}
\Phi_{ER}=\frac{\Phi_{R}F^2+\Phi_{Z}F_{Z}^2+2FF_Z(\theta_{x+} + \theta_{y-}+2\pi)}{F^2_{ER}}.
\label{eq:PhiER}
\end{equation}
If, as an example, we adopt the typical values $F=F_Z$, $\theta_{x+} = \theta_{y-}=\pi/2$, and $\Phi_R=\pi$, then
\begin{equation}
\Phi_{ER}=\frac{\Phi_{Z}+7\pi}{4},
\end{equation}
so that the net twist is roughly a quarter of the initial twist plus one turn, which is what one would guess qualitatively from Figure \ref{fig5}.

\section{3D Zipper  Reconnection Phase}
\label{sect_4}

We assume that the zipper reconnection phase is composed of multiple individual reconnection events between tube pairs of equal flux.  When the initial state is a sheared arcade these elementary events are simple zippettes (Figure \ref{fig2}) as analysed in Section \ref{sect_3.1}, whereas when the initial state includes a flux rope they are helical zippettes (Figure \ref{fig5}) as described in Section \ref{sect_3.3}.  

\subsection{\bf Zipper Reconnection in a Sheared Arcade}
\label{sect_4.1}

\begin{figure}[h]
{\centering
 \includegraphics[width=12cm]{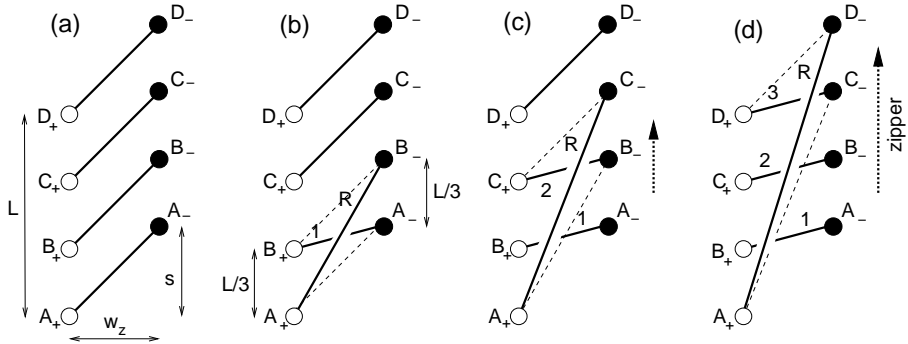}
\caption{A sequence of zippettes for zipper reconnection in a sheared arcade consisting of   $N=4$ flux tubes.  In the initial state (a) all flux tubes (thick solid lines) are parallel, connecting sources $A_+\to A_-$, $etc$.  The reconnection sequence consists of 3 individual reconnection zippettes (b)--(d).  The tubes just eliminated by reconnection are depicted by thin dashed lines.  The overlying twisted flux tube is designated $R$, and the reconnected arcade tube generated in reconnection event $n$ is designated with that number, \textit{i.e.} 1, 2, or 3.}
\label{fig6}}
\end{figure}

The zipper reconnection occurs between two rows of $N$ identical, equally-spaced sources arranged along parallel lines of length $L$, separated by $w=w_z$.  The $N$ positive sources are denoted by $A_+$, $B_+$, $C_+$, $etc.$, and the $N$ negative sources, $A_-$, $B_-$, $C_-$, $etc.$, as depicted in Figure\ \ref{fig6}.   The initial arcade is sheared by $\bm\bar{s}=s/w_z$ by displacing the entire row of negative sources northward a distance $s$.  For illustration Figure\ \ref{fig6} shows the $N=4$ case with $s>0$, but our calculation is equally valid when $s<0$.    Both lines of $N$ sources extend a distance $L$, and are therefore spaced by $L/(N-1)$.  The initial state is an arcade of $N$  flux tubes, $A_+A_-$, $B_+B_-$, $C_+C_-$, $etc$.  We assume  all have the same initial twist, $\Phi_0$: for the most part, we naturally assume $\Phi_0=0$ for our basic analysis, but we include it here for completeness.

The reconnection occurs in a sequence of individual zippette events.  The $n^{\rm th}$ zippette  produces new flux tubes with identical twist (thanks to helicity equipartition), which we designate $\Phi_n$.  We begin by considering a sequence whose first event is at the southern (lower) end, between $A_+A_-$ and $B_+B_-$, producing an overlying flux rope $A_+B_-$ and a twisted arcade tube $B_+A_-$; these are designated $R$ and $1$ in Figure\ \ref{fig6}b, and each has twist $\Phi_1$.   Each subsequent event reconnects the twisted flux rope anchored at $A_+$, denoted $R$ in Figure\ \ref{fig6}, with an unreconnected tube in the arcade.  In this way the reconnection spreads along the arcade with its right-hand footpoint sweeping northward along the negative polarity line like a zipper. It  terminates after $N-1$ events, leaving an overlying flux rope connecting the southern-most positive source ($A_+$) to the northern-most negative source ($D_-$ in Figure\ \ref{fig6}). The arcade has thus been reduced in flux by a factor $(N-1)/N$. In the case shown, the arcade's shear has been reduced, but, if $s$ were negative, the process would have {\em increased} the magnitude of the arcade's shear.

We apply the general formulation from Section \ref{sect_3.1} to the first reconnection zippette by associating the generic tube $X_+X_-$ with $A_+A_-$ and $Y_+Y_-$ with $B_+B_-$.  Those assignments lead to 
 $\bm\bar{t}=\bm\bar{s}=s/w_z$, $\bm\bar{r}=\bm\bar{\ell}=\bm\bar{L}/3$, and thus $\bm\hat{t}=\bm\bar{s}$, for the case $N=4$ shown in Figure\ \ref{fig6}.  Using these assignments in the general expression (\ref{eq:dPhi2}) yields the first twist increment as
\begin{equation}
  \tan(\Delta\Phi_1^{\rm m}) ~=~ -{2\bm\bar{L}^2\bm\bar{s}\over 9(\bm\bar{s}^2+1)^2 - (\bm\bar{s}^2-1)\bm\bar{L}^2} ~.
  	\label{eq:dPhi_g1}
\end{equation}
Since both initial flux tubes have initial flux $\Phi_X=\Phi_Y=\Phi_0$, the twist in the overlying flux rope ($A_+B_-$) is 
\begin{equation}
  \Phi_1 ~=~\Phi_0~+~\Delta\Phi_1^{\rm m}~. 
  	\label{eq:Phi_1} 
\end{equation}
This is also the twist in region $L$, which is $B_+A_-$.  Since that region reconnects no further, that twist value  remains in that flux tube. 
It can be shown from Equation (\ref{eq:dPhi_g1}) that, when $\bm\bar{s}>0$ then $0<\Delta\Phi_1^{\rm m}<\pi$, whereas when $\bm\bar{s}<0$ then $\pi<\Delta\Phi_1^{\rm m}<2\pi$ (see Appendix).

The second reconnection zippette occurs between the overlying flux rope $A_+B_-$ with twist $\Phi_1$ and the next arcade tube in line $C_+C_-$ (see Figure\ \ref{fig6}c).  We apply the generic result by associating $X_+X_-$ with tube $A_+B_-$ and $Y_+Y_-$ with $C_+C_-$.  The generic variables then take the values $\bm\bar{\ell}=2\bm\bar{L}/3$, $\bm\bar{r}=\bm\bar{L}/3$, 
$\bm\bar{t}=\bm\bar{s}+\bm\bar{L}/3$, and therefore $\bm\hat{t}=\bm\bar{s}$.  Making these substitutions in Equation (\ref{eq:dPhi2}) gives
\begin{equation}
  \tan(\Delta\Phi_2^{\rm m}) = -{2\bm\bar{L}^2(6\bm\bar{s}+\bm\bar{L})\over (9\bm\bar{s}^2 + 3\bm\bar{L}\bm\bar{s} - 2\bm\bar{L}^2 - 9)\, ( 3\bm\bar{s}^2 + \bm\bar{L}\bm\bar{s} - 3\bigr) 
  +3(6\bm\bar{s}+\bm\bar{L})^2 } ~.
  	\label{eq:dPhi_g2}
\end{equation}
Since flux tube $C_+C_-$ has twist $\Phi_Y=\Phi_0$, the twist in the new overlying flux tube, $A_+C_-$, is
\begin{equation}
  \Phi_2 ~=~ \hbox{${1\over2}$}(\Phi_0+\Phi_1) ~+~ \Delta\Phi_2^{\rm m} ~=~ \Phi_0 ~+~ \half\Delta\Phi_1^{\rm m} ~+~ \Delta\Phi_2^{\rm m} ~,
  	\label{eq:Phi_2}
\end{equation}
where the final expression results after substituting from Equation  (\ref{eq:Phi_1}).  The self-helicity from the first-generation flux rope ($\Phi_1$) is equally divided between the two tubes produced in the reconnection, designated $R$ and $2$ in Figure\ \ref{fig6}c.  Only half of the reconnection-created twist ($\Delta\Phi^{\rm m}_1$) ends up in the second-generation flux rope.

\begin{figure}[h]
{\centering
 \includegraphics[width=12cm]{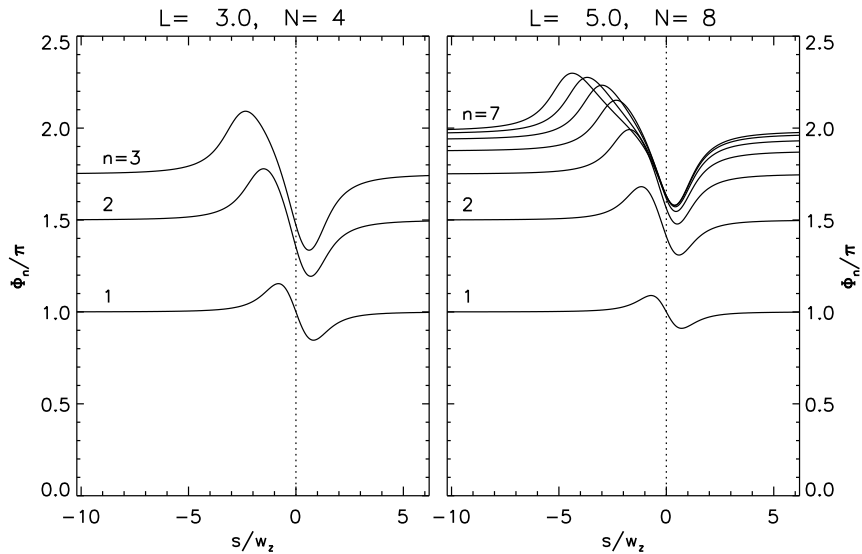}
\caption{Twist values in each of the newly reconnected sections, and in the final flux rope, for the zipper phases as a function of shear $\bm\bar{s}=s/w_z$ for two different cases, both involving initially untwisted flux ropes ($\Phi_0=0$).  Left: an arcade of length $L=3w_z$, resolved into $N=4$ components.  Right: an arcade of length $L=5w_z$, resolved into $N=8$ components.}
\label{fig7}}
\end{figure}

Repeating this procedure one more time yields the twist $\Phi_3$ in the final overlying flux tube $A_+D_-$.  The left panel in Figure\ \ref{fig7} plots this value, as well as $\Phi_1$ and $\Phi_2$, for the case $\bm\bar{L}=3$, $\Phi_0=0$, for a range of values of $s$, both positive and negative.  Note that the twist in the rope increases with each reconnection event.  This is natural since each reconnection event introduces a new crossing into the configuration and thus contributes a typical twist of
 $\Delta\Phi^{\rm m}_n\approx\pi$
(although it can lie between 0 and $2\pi$, see Appendix).
 This does not, however, lead to a final twist $\Phi_3\approx 3\pi$, since at each stage the flux rope loses half its accumulated twist to the newly created arcade loop.

In Figure \ref{fig7}, the graph of the rope twist ($\Phi_1$) after the first zippette from Equation (\ref{eq:dPhi_g1}) is antisymmetric about $s=0$ relative to $\pi$. It shows a twist of $\pi$ for $s=0$ and a twist that is larger for $s<0$ but smaller for $s>0$ and that tends to $\pi$ as $|s|$ tends to infinity. This may be understood from the symmetric location of the sources when $s=0$, since conservation and equipartition of magnetic helicity then imply that the reconnection adds exactly half a turn of twist. 
For $s=0$, Figure \ref{fig6} has $A_-$, $B_-$, $C_-$ and $D_-$ lying directly opposite $A_+$, $B_+$, $C_+$ and $D_+$, so that the tangent of the first twist decrement ($\tan \Delta \Phi_1^{\rm m}$) vanishes from Equation (\ref{eq:dPhi_g1}) and $\Delta \Phi_1^{\rm m}=\pi$. When $s<0$ the negative sources lie below their partners and the overlying reconnected rope $A_+B_-$ is shorter than when $s>0$. This results in greater twist when the shear is negative than for an equivalent positive shear. Furthermore, the antisymmetric property (evident in Equation (\ref{eq:dPhi_g1})) arises from the antisymmetric nature of the setup for the first zippette. 

However, the asymmetry is lost after the second and subsequent reconnections, where $\tan\Delta\Phi_n^{\rm m}\neq 0$ when $s=0$ and $\Phi_2$ tends to $3\pi/2$ as $|s|$ tends to infinity.  This can be seen from Equation (\ref{eq:dPhi_g2}) and Figure \ref{fig7}. It arises from a comparison of the geometry of Figure \ref{fig6} when $s<0$ and $s>0$. Thus, since the reconnection is proceeding in the direction from $B_-$ to $C_-$ to $D_-$, after the second reconnection the geometrical location of the reconnected rope is no longer antisymmetric.

The same procedure can be applied to cases with more flux tubes in the initial arcade.  The twist in the rope after its 
$n^{\rm th}$ zippette is 
\begin{equation}
  \Phi_n ~=~  \hbox{${1\over2}$}\Phi_{0} ~+~ \hbox{${1\over2}$}\Phi_{n-1} ~+~ \Delta\Phi_n^{\rm m} ~~.
  	\label{eq:Phin}
\end{equation}
The right panel of Figure\ \ref{fig7} shows all twist values for the case with $N=8$, $\bm\bar{L}=5$, and $\Phi_0=0$. It appears that, for cases when $|\bm\bar{s}|\gg\bm\bar{L}$, this approaches an upper bound of $\Phi_n\approx 2\pi$.  This is the asymptotic fixed point for relation (\ref{eq:Phin}) when $\Delta\Phi^{\rm m}_n\approx\pi$.

\subsection{\bf Zipper Reconnection in an Arcade Overlying a Flux  Rope}
\label{sect_4.2}

\begin{figure}[h]
{\centering
 \includegraphics[width=12cm]{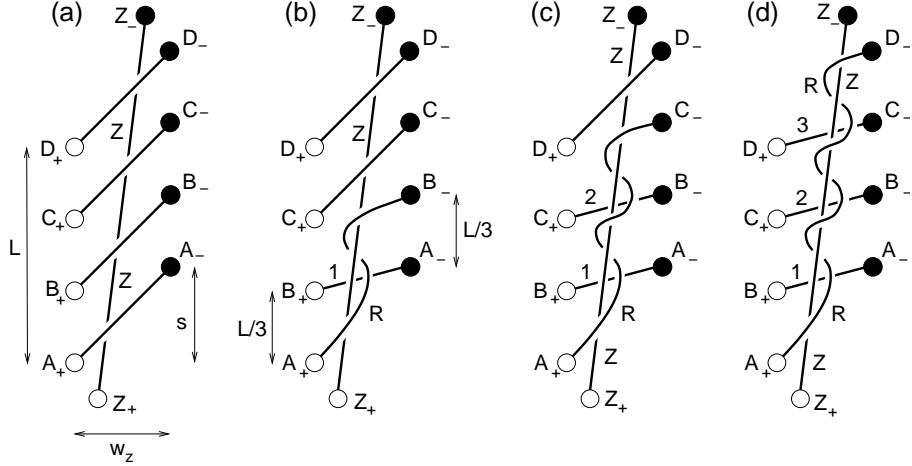}
\caption{A sequence of helical zippettes for zipper reconnection in an arcade containing an initial flux rope and consisting of $N=4$ flux tubes.  In the initial state (a) all flux tubes (thick solid lines) are parallel, connecting sources $A_+\to A_-$, etc.  These overlie the pre-existing flux rope $Z$ connecting $Z_+\to Z_-$. The reconnection sequence consists of 3 individual reconnection events (b)--(d).  The overlying twisted flux tube is designated $R$, and the reconnected arcade tube generated in reconnection event $n$ is designated with that number, \textit{i.e.} 1, 2, or 3.}
\label{fig8}}
\end{figure}

Consider next zipper reconnection in an arcade that initially overlies a flux rope of twist $\Phi_r$, say (Figure \ref{fig8}). In this case it occurs by a series of helical zippettes as described in Section \ref{sect_3.3}, but the surprising result there was that the initial flux rope does not affect the twists or helicity of the tubes that participate in the reconnection. Thus, the twist in the core of the resulting flux rope is the same as in the initial flux rope ($\Phi_r$), and it is surrounded by a sheath with the twist that is calculated in Figure \ref{fig7}.

Note that, after the first zippette, each of the $n-1$ subsequent zippette reconnections adds no flux to the new flux rope $R$ but adds an extra turn of twist
to the composite structure. Thus, suppose that, as in Section \ref{sect_3.3}, we combine the new flux rope ($R$) and the initial one ($Z$) to give an erupting flux rope of flux
\begin{equation}
F_{ER}=F+F_Z
\label{eq_FER}
\end{equation}
and twist $\Phi_{ER}$, which is determined by the fact that its self-helicity is the sum of the self- and mutual helicities of $R$ and $Z$.  The resulting twist is of the same form as Equation (\ref{eq:PhiER}), except that the presence of the $n$ turns implies that the term $2\pi$ in the last bracket is replaced by $2n\pi$, and also the angles $\theta_{x+}$ and $\theta_{y-}$ now refer to angles $Z_-A_+Z_+$ and $Z_-D_-Z_+$ in Figure \ref{fig8}. Thus, the net twist becomes
\begin{equation}
\Phi_{ER}=\frac{\Phi_{R}F^2+\Phi_{Z}F_{Z}^2+2FF_Z(\theta_{x+} + \theta_{y-}+2n\pi)}{F^2_{ER}}.
\label{eq_PhiER}
\end{equation}

If, as an example, we adopt the typical values $F=F_Z$, $\theta_{x+} = \theta_{y-}=\pi/2$, and $\Phi_R=2\pi$, then
\begin{equation}
\Phi_{ER}=\frac{\Phi_{Z}}{4}+(n+1)\pi,
\end{equation}
so that the net twist is roughly a quarter of the initial twist plus $(n+1)/2$ turns. 

The number of turns is determined  by the number of zippettes, which depends on  how far the ribbons extend along the polarity inversion line and so how much magnetic flux is contained within them. In other words,
\begin{equation}
n=\frac{F_{\rm ribbon}}{F_R}, 
\label{number turns}
\end{equation}
where $F_R$ is the magnetic flux of the new flux rope ($A_+$ in Figure \ref{fig8}) and $F_{\rm ribbon}$ is the magnetic flux in one of the flare ribbons when it is first fully formed ($A_+B_+C_+D_+$ in Figure \ref{fig8}).
The greater the magnetic flux in the initial flare ribbons, the larger the number of turns and so the more highly twisted is the core of the erupting flux rope.

\subsection{\bf Other Sequences of Zipper Reconnection}
\label{sect_4.3}

\begin{figure}[h]
{\centering
 \includegraphics[width=10cm]{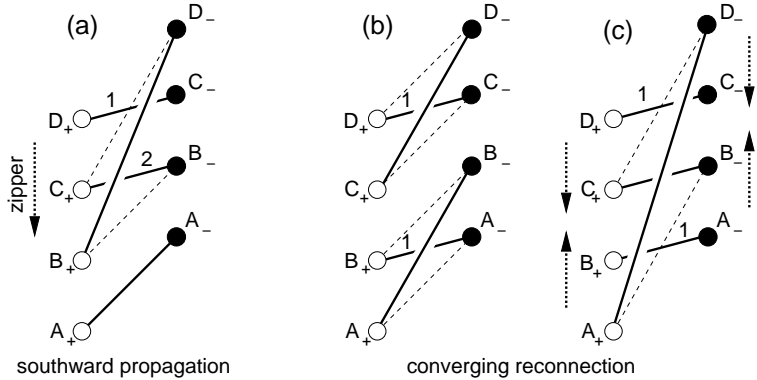}
\caption{Two sequences of zipping reconnection with $N=4$ flux tubes that are different from the case of northward propagation depicted in Figure\ \ref{fig6}.  (a) Southward propagation  produces similar results but with twist distributed in the opposite order, so that $D_+C_-$ has twist $\Phi_1$ rather than $\Phi_3$. (b) In a converging sequence, the third reconnection event (shown in (c)) occurs between two flux ropes ($A_+B_-$ and $C_+D_-$) with twist $\Phi_1$.}
\label{fig9}}
\end{figure}

The previous two subsections  consider a particular reconnection sequence in which reconnection propagates northward to produce an overlying flux rope linking the southern most positive source ($A_+$) to the northern most negative source ($D_-$).  This same final connection can be produced by other reconnection sequences, whether there is an initial flux rope present or not.  The mutual helicity of the final configuration is determined by connectivity, so the change in mutual helicity is independent of the sequence that produces it.   Since total helicity is conserved, the self-helicity is also  independent of the sequence, so it is tempting to conclude that the final state itself is independent of the sequence.  The distribution of self-helicity depends, however, on the sequence owing to our assumption of helicity equipartition.  We therefore find different distributions of twist for different reconnection sequences.  

To illustrate this sequence-dependence, consider a second sequence where reconnection propagates southward beginning from two northern-most tubes, $C_+C_-$ and $D_+D_-$.  This event is identical to the first event in the original sequence, so $\Phi_1$ is the same.  The resulting overlying flux rope ($C_+D_-$) reconnects next with the untwisted tube to its south ($B_+B_-$), as shown in Figure\ \ref{fig9}a.   This event is not identical with the second event from before, since $\bm\bar{t}=\bm\bar{s}$, $\bm\bar{l}=\bm\bar{L}/3$, $\bm\bar{r}=2\bm\bar{L}/3$, and $\bm\hat{t}=\bm\bar{s}+\bm\bar{L}/3$.  
It is an inverted version of that event, and can be transformed into it by swapping $\bm\bar{\ell}$ and $\bm\bar{r}$ and swapping $\bm\bar{t}$ and $\bm\hat{t}$.  It is evident from the symmetric form of Equation  (\ref{eq:dPhi2}), that this transformation leaves $\Delta\Phi^{\rm m}_2$ unchanged.  This means that $\Phi_2$ is the same, and similar reasoning applies to $\Phi_3$.  We therefore see that southward-propagating reconnection leads to an overlying flux rope and a set of arcade tubes with the same twist as in the northward-propagating case.  These tubes are, however, arranged in the opposite order, with the most-twisted tube ($3$) located at the south end rather than the north end as it was before.

The foregoing argument can be applied more generally to any sequence in which a single overlying flux rope reconnects with the neighboring unreconnected tube on either side --- north or south.  Due to symmetry, $\Delta\Phi^{\rm m}_n$ is the same regardless of the direction in which the reconnection proceeds.  This is even true for a sequence where northward and southward propagation is interleaved.  

The results are not the same, however, if multiple overlying ropes are produced through separate sequences and then merged.  An example of such a case, depicted in Figure\ \ref{fig9}b--c, has two flux ropes, each with twist 
$\Phi_1$, reconnecting to form the final overlying rope.  That final reconnection event is characterized by $\bm\bar{t}=\bm\hat{t}=\bm\bar{s}+\bm\bar{L}/3$ and $\bm\bar{\ell}=\bm\bar{r}=2\bm\bar{L}/3$, from which Equation 
(\ref{eq:dPhi2}) yields
\begin{equation}
  \tan(\Delta\Phi^{\rm m}_{2'}) = -{24\bm\bar{L}^2(3\bm\bar{s}+\bm\bar{L})\over (9\bm\bar{s}^2 + 6\bm\bar{L}\bm\bar{s} +\bm\bar{L}^2 +9)^2 - 4\bm\bar{L}^2(9\bm\bar{s}^2 + 
  6\bm\bar{L}\bm\bar{s} +\bm\bar{L}^2 - 9 )} ~,
    	\label{eq:dPhi_g2p}
\end{equation}
which is different from Equation  (\ref{eq:dPhi_g2}) for the simple northward case.  Moreover, since the reconnecting tubes have identical twist, the products have twist
$\Phi_{2'}=\Phi_1+\Delta\Phi^{\rm m}_{2'}$.  This is in general greater than the $\Phi_2$ produced by simple northward propagation.  Moreover, in order that total helicity is the same in both scenarios, we find the relation
\[
  2\Phi_{2'}+2\Phi_1 ~=~ \Phi_1+\Phi_2+2\Phi_3 ~.
\]
This can be rearranged to find $\Phi_{2'}=\Phi_3+\Delta\Phi_2^{\rm m}/2-\Delta\Phi^{\rm m}_1/4$, so the final flux rope is more twisted in this merging scenario than it is in the case where a single flux rope is progressively formed.

Reconnection sequences producing the opposite sense of connectivity naturally produce flux ropes with the opposite twist.  
Thus, for example, consider the basic simple zippette process in Section \ref{sect_3.1} and suppose for simplicity that $\Phi_X=\Phi_Y=0$, so that Equation (\ref{eqnconshel}) becomes
\begin{equation}
 {F^2\over\pi}\Phi_R =H^{\rm m}_{XY} ~-~ H^{\rm m}_{UR},
\end{equation}
which implies that the twist in the rope due to the change in mutual helicity is
\begin{equation}
  \Phi_R ~=~ \theta_3+\theta_4+\theta_2-\theta_1.
  \label{eqnPhiR}
\end{equation}
Thus, $\Phi_R $ is exactly equal to $\pi$ in the special case of a rectangle (when $\theta_3=\theta_4=\pi/2$ and $\theta_2=\theta_1$).
For the configuration shown in Figure \ref{fig2}, the twist is positive since the initial mutual helicity ($H^{\rm m}_{UR}$) is positive and the final mutual helicity ($H^{\rm m}_{XY}$) of the flux tube $R$ lying over the tube $U$ is negative. This may be seen from the right-hand rule in the sense that, if the fingers of the right hand are directed along the overlying magnetic field, then the sign is positive if the underlying magnetic field is in the direction of the thumb.

If the tube $R$ instead passes under the tube $U$, the helicity $H^{\rm m}_{UR}$ becomes positive and $\theta_3+\theta_4$ is replaced by $\theta_5+\theta_6$,  where $\theta_5$ and $\theta_6$ are the angles $X_+Y_+Y_-$ and $X_+X_+Y_-$.  However, $\theta_3+\theta_4+\theta_5+\theta_6=2\pi$, so the net effect is that $\theta_3+\theta_4$ is replaced by $\theta_3+\theta_4-2\pi$ in Equation (\ref{eqnPhiR}), which now implies that $\Phi_R<0$ and for a rectangle $\Phi_R=-\pi$. (The result $\Phi_R<0$ arises because in Figure \ref{fig2} the angle $X_+Y_+X_-$ is $\theta_2$ since $X_+Y_+$ and $X_-Y_-$ are parallel and so $\theta_3$ and $\theta_2$ form two angles of the triangle $X_+Y_+X_-$. Thus, $\theta_3+\theta_2<\pi$, $\theta_4<\pi$ and $\theta_1>0$,  so that $\theta_3+\theta_2+\theta_4-\theta_1<2\pi$, as required.)

Consider also a version of northward propagation (\textit{i.e.},\ Figure\ \ref{fig6}), but where the negative foot ($A_-$) remains fixed at each stage, and the reconnection propagates along the positive sources until it reaches $D_+$.  This is a mirror image of the original case depicted in Figure\ \ref{fig6}a.  It can be converted to that same sequence by taking $x\to-x$, $s\to-s$, and reversing polarities, $B_z\to-B_z$.  The act of spatial reversal ($x\to-x$) changes the sign of the helicity.  The set of twist variables for this reversed case is related to the original by $\Phi_n^{(r)}(s)=-\Phi_n(-s)$, where the functions 
$\Phi_n(s)$ are plotted in Figure\ \ref{fig7}a.

\section{The Quasi-2D Main-Phase Reconnection Process} 
\label{sect_5}

We next assume that a twisted flux rope ($Z$), which is produced (or enhanced if a flux rope is present initially) through zipper reconnection of an arcade, erupts.  In so doing it pushes the overlying flux ahead of it.  This flux drapes around the erupting rope forming a current sheet beneath it, often called the ``flare current sheet" \cite{priest00,karpen12,longcope14}.  Reconnection at this trailing current sheet disconnects some of the overlying flux from the photosphere, thereby assisting in the eruption of flux rope $Z$.  In strictly two-dimensional models this so-called ``flare reconnection'' is a case of self-reconnection, since field lines reconnect with themselves.  The result is a closed underlying arcade loop and a disconnected loop completely encircling flux rope $Z$.  

In quasi-two-dimensional or three-dimensional scenarios, however, the reconnection involves two different field lines, connecting two pairs of footpoints, say $X_+\to X_-$ and $Y_+\to Y_-$,
as has been mentioned in some simulations of eruptive flares and CMEs \cite{manchester04,fan07a} and made explicit in Figure 5 of \inlinecite{aulanier12}.
  The reconnection thus creates two closed field lines and no disconnected loop results.  One of these field lines ($U$) is underneath rope $Z$, while the disconnected loop in the two-dimensional picture becomes the other field line ($R$), which remains above $Z$ and twists about it. This is in fact the scenario we designated a helical zippette, as discussed in Section \ref{sect_3.3} and illustrated in Figure \ref{fig5}. This phase is ``quasi-two-dimensional'' in the sense that it produces a rising arcade of flare ribbons and two separating H$\alpha$ ribbons as a series of nested flux sheaths reconnect.

\subsection{\bf Reconnecting the First Overlying Sheath of Flux by Helical Zippettes}
\label{sect_5.1}

\begin{figure}[h]
{\centering
 \includegraphics[width=12cm]{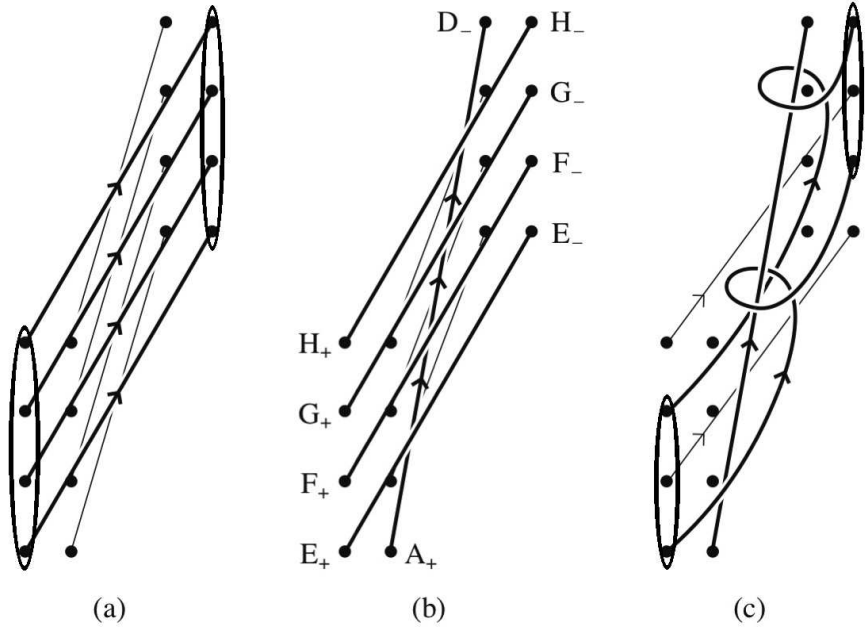}
\caption{The two inner-most sets of field lines of the coronal arcade seen from above, showing: (a) the initial sheared state; (b) the situation after the zipper phase of the flare when reconnection has proceeded parallel to the polarity inversion line to produce the zipper flux rope ($A_+D_-$) lying under the next set of field lines joining footpoints $E_+E_-,\ F_+F_-,\ G_+G_-,\ H_+H_-$; (c) the first part of the main phase after reconnection has progressed sideways to reconnect the next sheath of field lines and create a spiral sheath that wraps around the zipper flux rope and enhances its flux and magnetic helicity.}
\label{fig10}}
\end{figure}
Consider first the main phase of the eruption of a sheared arcade containing a flux rope created by the zipper phase. This phase  consists of multiple helical zippettes of the kind  analyzed in Section \ref{sect_3.3}.  Figure \ref{fig10} illustrates how this might proceed for the next layer outside the zipper flux rope in the $N=4$ case used for illustration above.  That next layer consists initially of 4 parallel flux ropes, $E_+E_-$, $F_+F_-$ $etc.$, with identical initial twist $\Phi_0$ which make up a sheath of flux whose feet form the narrow ellipses indicated in Figure \ref{fig10}a. (We assume $\Phi_0$ is constant here, but later consider the possibility that it varies between layers.)  The eruption of the twisted flux rope ($A_+D_-$) formed in the zipper phase, leads to reconnection around that erupting tube.  

Figure \ref{fig10} shows the case where the basic process in that reconnection occurs in two separate events like the $XYZ\to URZ$ case described in Section \ref{sect_3.3}.  The result is two underlying arcade loops, $F_+E_-$ and $H_+G_-$ which form an underlying sheath of flux, and two flux ropes, $E_+F_-$ and $G_+H_-$; the latter form a sheath of flux, which wraps around the flux rope $A_+D_-$ and which has feet forming two narrow ellipses shown in Figure \ref{fig10}c.  All flux tubes have the same internal twist, $\Phi_1$.  This is the same as in Section \ref{sect_3.3}, given by Equation  (\ref{eq:dPhi_g1}) but with slightly different values of $\bm\bar{s}$ and $\bm\bar{L}$.   Figure \ref{fig1} shows that $s$ and $L$ are the same as in the zipper phase, but $w>w_z$, so the re-scaled variables $\bm\bar{s}=s/w$ and $\bm\bar{L}=L/w$ are smaller at this phase; they are smaller still at each successive phase thereafter.  Given the complex structure of the curves in Figure\ \ref{fig10}, we need to incorporate the structure of the arcade in order to determine in which sense $\Phi_1$  changes as a result.

\begin{figure}[h]
{\centering
 \includegraphics[width=10cm]{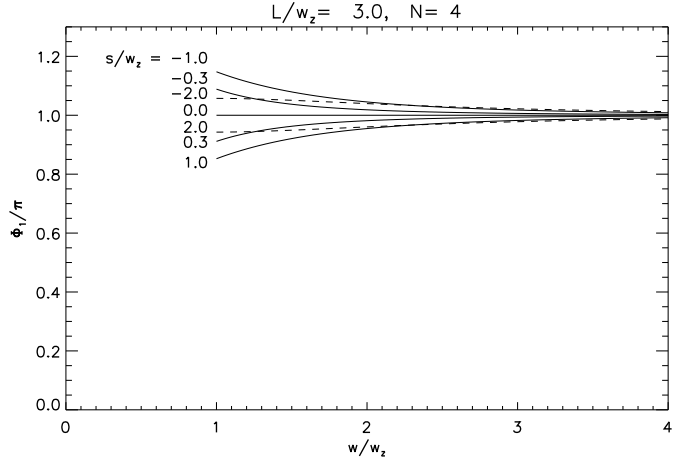}
\caption{Plot of $\Phi_1$  against $w$ for values of $s/w_z=-2,\,-1,\,-0.3,\,0,\,0.3,\,1,\,2$.  The extreme values, $s/w_z=\pm2$, are plotted with dashed curves from clarity.  
The other parameters, $L/w_z=3$ and $N=4$ are the same as for Figure\ \ref{fig7}a.  It is evident from that plot that the twist is maximum and minimum when $s=-1$ and $s=+1$ respectively.}
\label{fig:fig11}}
\end{figure}
The twist is determined by the zipper process inside the core of the arcade ($w<w_z$) and by the main phase process outside the core ($w_z<w<w_f$), where $\Phi_1$ is given by Equation  (\ref{eq:Phi_1}) with ${\bm\bar s}=s/w$. The resulting graph of twist as a function of $w$ for $\Phi_0=0$ and several values of $s$ is shown in Figure\ \ref{fig:fig11}.  The internal twist is only a weak function of $w$, and generally increases for positively sheared arcades ($s>0$).  This internal twist does not, however, account for the fact that the flux from the second layer ($E_+$--$H_-$) ends up wrapped around the central core.  We account for that below.

As in our discussion of a helical zippette in Section \ref{sect_3.3}, we may combine the main phase flux rope (of flux $F_M$, and twist $\Phi_M$, say) and the zipper flux rope (of flux $F_{ER}$ and twist $\Phi_{ER}$ from Equations (\ref{eq_FER}) and (\ref{eq_PhiER})) to give a total erupting flux rope of flux 
\begin{equation}
F_{T}=F_M+F_{ER}
\end{equation}
and twist $\Phi_{T}$, which is determined by the fact that its self-helicity should be the sum of the self- and mutual helicities of the main phase and zipper flux.
The resulting twist has the same form as Equation (\ref{eq:PhiER}):
\begin{equation}
\Phi_{T}=\frac{\Phi_{M}F_M^2+\Phi_{ER}F_{ER}^2+2F_MF_{ER}(\theta_{x+} + \theta_{y-}+2\pi)}{F^2_{T}}.
\end{equation}
Thus, the twist depends crucially on the zipper twist and the ratio of the main phase and zipper fluxes, but for a weak zipper twist it is typically one turn.

\subsection{\bf Some Ways of Increasing the Twist}
\label{sect_5.2}

The simple reconnection scenario described in Section \ref{sect_5.1} starting with a sheared arcade results in a flux rope which is roughly uniformly twisted by about one turn surrounding a less twisted core.  This is a result of the main phase reconnection occurring around a core created by the initial zipping reconnection. The latter has been found to produce no more twist than approximately one turn.  

Remembering that the interplanetary observations of magnetic clouds \cite{hu14} suggest either a flux rope with a constant twist of 1.5 to 3 turns or one with an enhanced core twist of up to 5 turns, we now consider extra effects that may increase the twist above that of our basic cases. 

The most natural way is to adopt as our initial state a coronal arcade that overlies a pre-formed flux rope with large twist $\Phi_r$, as suggested in Section \ref{sect_4.2}.  The zipping phase would then be by helical zippettes, which produces a central concentration in $\Phi_r$, around which there can be many turns.

In Section \ref{sect_4.3} we have also shown how zipper reconnection occurring in several places along the polarity inversion line (rather than starting at one end) increases twist above the unidirectional value.

Another possibility is to allow each arcade flux tube to possess its own initial internal twist $\Phi_0$. This adds a constant to the twist at each generation.  Such a pattern is 
evident in the first two generations, given by Equations (\ref{eq:Phi_1}) and (\ref{eq:Phi_2}).  Induction using Equation (\ref{eq:Phin}) shows that the pattern persists to arbitrary $n$.  Any non-vanishing initial twist is thus added to every curve in Figure\ \ref{fig7}.  If we start initially with a sheared arcade, such addition could twist the reconnected flux rope $R$ to an angle significantly above $2\pi$ to trigger a kink instability \cite{hood79a}.

A further possibility  during the main phase (Section \ref{sect_5.1}) considered in detail in the next subsection is to allow extra reconnection along each sheath between the flux tubes that make up the sheath  in Figure \ref{fig10}. 

\subsection{\bf 3D Main Phase Reconnection -- Extra Reconnection Within Each Sheath Along Its Length}
\label{sect_5.3}
After the quasi-2D main phase process, two overlying flux tubes ($E_+F_-$ and $G_+H_-$) in Figure \ref{fig10}c could reconnect to produce another underlying arcade tube ($G_+F_-$) and a single overlying flux tube ($E_+H_-$).  This event consists of flux tubes $X$ and $Y$ each initially wrapped $N_x=1$ and $N_y=1$ times around the central tube, $Z$.  Following the previous discussion, the $R$ tube wraps $N_x+N_y+1=3$ times around the erupting rope.  Otherwise the sequence is the same as that for converging reconnection illustrated in Figure\ \ref{fig9}b--c, so the final arcade loop, and the wrapping tube, are twisted by $\Phi_{2'}$, as found in Section \ref{sect_4.3}.

This produces an erupting flux rope composed of two separate strands.  The inner core is tube $A_+D_-$ produced by the zipper phase, with internal twist $\Phi_3$.  Wrapped about this is a second strand ($E_+H_-$) with internal twist $\Phi_{2'}$.  We can lump these components into a single erupting flux rope with $F_{z,2}=2F$ whose footpoints lie midway between the feet of its components: \textit{i.e.},\ midway between $A_+$ and $E_+$ in the south and midway between $D_-$ and $H_-$ in the north.  The self-helicity of this tube is a sum of the self-helicities and mutual helicity of its constituents
\begin{equation}
  H^{\rm s}_{z,2} ~\approx~ {F^2\over2\pi}\Phi_3 ~+~ {F^2\over 2\pi}\Phi_{2'} ~+~ {F^2\over \pi}\Bigl(\, 2\theta_{r}+6\pi\,\Bigr) ~=~ {F_{z,2}^2\over2\pi}\bm\bar{\Phi}_{z,2} ~,
  	\label{eq:Phiz1}
\end{equation}
after using the footpoints of the composite tube in Equation  (\ref{eq:HmRZ}) with 
\begin{equation}
  \theta_{x+}~=~\theta_{y-}~=~ \theta_r ~=~\tan^{-1}\left[{\bm\bar{s}+\bm\bar{L}\over (w+w_z)/2}\right] ~.
\end{equation}
The effective twist of the erupting rope is 
\begin{equation}
  \bm\bar{\Phi}_{z,2} ~=~ {1\over4}\Bigl(\, \Phi_3 ~+~\Phi_{2'} ~+~ 12\pi~+~4\theta_r\,\Bigr) ~.
  	\label{eq:Phiz2}
\end{equation}

The reconnection of the sheath of flux results, therefore, in an erupting flux rope with roughly the same connectivity and some additional twist.  The sequence we use for illustration  moves the flare ribbons outward, with little evident tendency to move northward or southward.  

\begin{figure}[h]
{\centering
 \includegraphics[width=12cm]{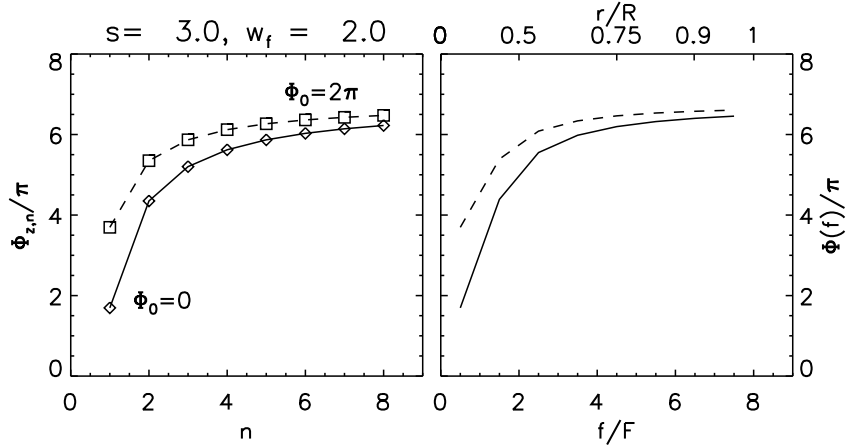}
\caption{The distribution of twist for a flux rope erupting from an arcade with $s=L=3w_z$.  Eight arcade layers run to $w=w_f=2w_z$, each represented by $N=4$  sources.
The left panel shows the mean twist $\bm\bar{\Phi}_{z,n}$ from Equation (\ref{eq:Phizn}) after the $n^{th}$ arcade layer for arcades with initial twist, $\Phi_0=0$ (diamonds) and $\Phi_0=2\pi$ (squares).  The right panel shows the distribution of twist, $\Phi(f)$, defined in Equation (\ref{eq:phif}) for the same two cases, as a function of flux $f$ or (top axis) radius $r/R$, where $R$ is the outer radius.}
\label{fig12}}
\end{figure}
The next layer  reconnects in a manner exactly like the previous one except that the central flux rope ($Z$) now has flux $F_{z,2}=2F$.  After the complete reconnection of  that second layer the central rope is wrapped three times by an overlying flux rope, twisted by $\Phi_{2'}$.  The flux of the  central rope is raised to $F_{z,3}=3F$, and its self-helicity is increased.    
Generalizing Equation  (\ref{eq:Phiz1}) to the $n^{\rm th}$ stage of reconnection gives
\begin{equation}
  {F_{z,n-1}^2\over2\pi}\bm\bar{\Phi}_{z,n-1} ~+~ {F^2\over 2\pi}\Phi_{2'} ~+~ 
 {F\,F_{z,n-1}\over \pi}\, (6\pi+2\theta_r)  ~=~ {F_{z,n}^2\over2\pi}\bm\bar{\Phi}_{z,n} ~,
\end{equation}
where $F_{z,n}=nF$ is the flux of the rope and $\bm\bar{\Phi}_{z,n}$ is its effective twist.   Such a requirement can be used to form a recursion relation for the effective twist:
\begin{equation}
  \bm\bar{\Phi}_{z,n} ~=~ {(n-1)^2\over n^2}\, \bm\bar{\Phi}_{z,n-1} ~+~{4(3\pi+\theta_r^{(n)})(n-1)\over n^2} ~+~ {1\over n^2}\, \Phi^{(n)}_{2'} ~ ,
  	\label{eq:Phizn}
\end{equation}
where we  introduce superscripts to $\Phi^{(n)}_{2'}$ and $\theta_r^{(n)}$ as a reminder that $\bm\bar{s}$ and $\bm\bar{L}$ decrease with increasing $n$.  The footpoint of the flux rope is now located at $(w_{n-1}+w_z)/2$ so
\begin{equation}
 \theta_r^{(n)} ~=~\tan^{-1}\left[{4(\bm\bar{s}+\bm\bar{L})\over 2w_n+w_{n-1}+w_z}\right] .
\end{equation}
Expression (\ref{eq:Phizn}) reduces to Equation  (\ref{eq:Phiz2}) for the case $n=2$ after defining $\bm\bar{\Phi}_{z,1}=\Phi_3$, as the result of the zipper phase. It also approaches the limit
\begin{equation}
  \bm\bar{\Phi}_{z,n} ~\to~ 6\pi ~+~2\theta_r^{(n)} ,~~n\gg1 ~,
  	\label{eq:Phi_lim}
\end{equation}
reflecting the dominant contribution of the wrapping of each flux tube added to the central rope.  Figure \ref{fig12} shows the mean twist values $\bm\bar{\Phi}_{z,n}$ for a flux rope with $L=s=3w_z$ and an arcade spanning $w_z<w<w_f=2w_z$.  The values are found by iterating Equation  (\ref{eq:Phizn}) for $n=1,\,2,\dots\, 8$.

The flux rope that ultimately erupts is built up by a sequence of reconnection phases, described above, which add both flux and magnetic helicity.  We designate the accumulated flux by
$f=F_{z,n}=nF$ and the effective twist within the central portion of flux $f$ by $\bm\bar{\Phi}(f)=\bm\bar{\Phi}_{z,n}$, which satisfies approximately the differential equation
\begin{equation}
  {{\rm d}\bm\bar{\Phi}\over {\rm d}f} \approx{\bm\bar{\Phi}_{z,n}-\bm\bar{\Phi}_{z,n-1}\over F} = {F-2f\over f^2}\bm\bar{\Phi} + {4(3\pi+\theta_r^{(n)})(f-F)\over f^2} 
  + {F\Phi_{2'}(f)\over f^2} ~.
  	\label{eq:Phi_bar}
\end{equation}
For an axisymmetric flux tube with twist density $\Phi(f)$, the mean twist within the central flux $f$ is the weighted average \cite{priest16}
\begin{equation}
  \bm\bar{\Phi}(f) ~=~ {2\over f^2}\int\limits_0^f\, \Phi(f')\, f'\, {\rm d}f' ~.
\end{equation}
Inverting this and using Equation  (\ref{eq:Phi_bar}) yields the twist density
\begin{eqnarray}
  \Phi(f) &=& {1\over 2f}{{\rm d}\over {\rm d}f}\Bigl[ \, f^2\, \bm\bar{\Phi}(f) \, \Bigr] \nonumber \\
  &=& 6\pi ~+~2\theta_r(f)~+~ {F\over 2f}\Bigl[\, \bm\bar{\Phi}(f)+\Phi_{2'}(f) - 12\pi - 4\theta_r(f)\,\Bigr] ~.
  	\label{eq:phif}
\end{eqnarray}
The result, plotted in the right panel of Figure\ \ref{fig12}, is close to a uniform twist of $\Phi\approx 6\pi$, due to the wrapping of flux from reconnecting the overlying layers.  A very small central core produced by the zipper phase is less twisted.

\begin{figure}[h]
{\centering
 \includegraphics[width=12cm]{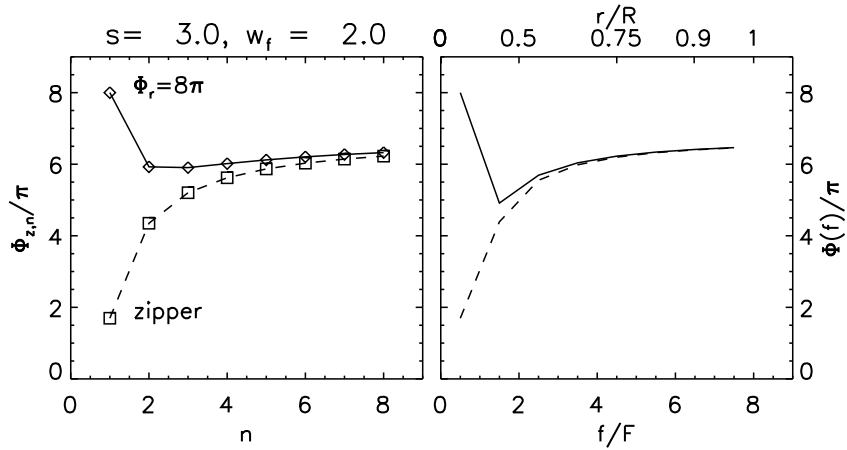}
\caption{The distribution of twist for an erupting flux rope beginning with a pre-formed flux rope with twist $\Phi_r=8\pi$.   The configuration is the same as in Figure \ref{fig12}, and the dashed curve, labeled ``zipper'', repeats the case $\Phi_0=0$ from that plot.}
\label{fig13}}
\end{figure}
The above scenario produces a flux rope which is roughly uniformly twisted surrounding a less twisted core.  A highly twisted core is instead produced if the initial state includes a pre-formed flux rope with large twist $\Phi_r$.  This assumes the role of flux rope $Z$ in Figure\ \ref{fig5}, and the main phase reconnection is of the form described in Section\ \ref{sect_4.2}.  The entire process  follows Equation  (\ref{eq:Phizn}), with $\bm\bar{\Phi}_{z,1}=\Phi_r$.  Figure \ref{fig13} shows the result when the initial flux rope has $\Phi_r=8\pi$.  This produces a central concentration in $\Phi(f)$, around which is relatively uniform twist closely matching  the previous case (see dashed curve).

Several obvious modifications are possible to this scenario or to the earlier zipper case.  The number of elements ($N$) in the outer layers may be increased.  We have seen in the previous section that increasing $N$ makes only a modest change to the flux rope produced by the zipper phase: $\Phi_{N-1}$ generally approaches $2\pi$ as $N$ increases.  It has a significant effect on the main phase, since it results in a flux rope wrapping $N-1$ times around the core.  The factor $3\pi+\theta_r^{(n)}$ is therefore replaced by $(N-1)\pi+\theta_r^{(n)}$ in Equation (\ref{eq:Phizn}).  More generally, the partitioning of layers could vary with distance, in which case the factor would be $(N_n-1)\pi+\theta_r^{(n)}$, and the profile would become less uniform and approaches $2(N_n-1)\pi+2\theta_r^{(n)}$, instead of Equation  (\ref{eq:Phi_lim}).  With this parametric freedom it appears possible to produce twist profiles with a wide variety of forms.  

\section{Discussion} 
\label{sect_6}

One of the key features of flares is that they start at some point along a polarity inversion line and then spread during the rise phase in a direction along it. Later, during the main phase the flare spreads outwards in a direction normal to the inversion line.  Another feature is that, when a flux rope that has originated in an erupting flare or a coronal mass ejection is observed in interplanetary space, it can have either a relatively uniform twist profile or a highly twisted core surrounded by a region where the twist is much more uniform.

We  present a simple model to try and explain these observations by adopting three assumptions, namely, conservation of magnetic flux and magnetic helicity and equipartition of magnetic helicity, which allow us to compare the pre-flare and flaring situation. We suggest that, during the phase of so-called 3D ``zipper reconnection",  reconnection spreads along the arcade away from the initiation site, and in so doing creates a twisted flux rope.  

This  either acts as the core for the erupting flux rope if the initial state is a sheared arcade,  or it  wraps around a  twisted flux rope that is present in the pre-flare state. 
In the former case a twisted flux rope is created with a moderate twist of typically only one turn.
In the latter case when a preflare flux rope is present,
 a much more highly twisted core can be produced with a typical twist of 
  \begin{equation}
\Phi_{ER}=\frac{\Phi_{Z}}{4}+(n+1)\pi,
\end{equation}
namely, roughly $(n+1)/2$ turns. 
Here
\begin{equation}
n=\frac{F_{\rm ribbon}}{F_R}, 
\end{equation}
where $F_R$ is the magnetic flux of the new flux rope  ($A_+$ in Figure \ref{fig8}) and $F_{\rm ribbon}$ is the magnetic flux in one of the flare ribbons when it is first fully formed  ($A_+B_+C_+D_+$ in Figure \ref{fig8}).
Hence the greater the number of times that the new flux rope reconnects with the  magnetic flux of the initial flare ribbons as it zippers its way along the polarity inversion line, the larger the number of turns produced.

The initial phase of a two-ribbon flare is clearly not produced by 2D reconnection, since the flare ribbons do not form instantaneously. Rather, the energy release is observed to be inhomogeneous and to fragment along the PIL: it starts at one location and then spreads along the PIL by what we term zipper reconnection. Thus, the magnetic flux is quantized, in the sense that only one part of the flux reconnects initially. We suggest that this initial quantum of flux then reconnects again with another quantum located further along the PIL and so the process continues in a zipper-like manner and forms the whole flare ribbon.  

The cause of this quantization may be that the initial resistive instability involves just one part of the whole configuration or is focused in one part with a certain quantum of flux; or perhaps in some events the photospheric flux itself is concentrated rather than being spread uniformly along the PIL. As we have seen the size of the quantum is important, since the resulting twist created by zipper reconnection is proportional to the ratio of the total flux to the quantum (see Equation (\ref{number turns})).

We have chosen to model the zipper process most simply by assuming the magnetic flux itself is fragmented in a series of flux sources located along the PIL ($e.g.$, Figure \ref{fig6}). It may be possible in future to simulate the process with an initial field that is not fragmented, but in which the magnetic reconnection begins at one location (as observed) rather than beginning simultaneously all along the PIL (as in a purely 2D model).

After the zipper phase, quasi-2D ``main phase reconnection" causes the reconnection to spread in a direction normal to the polarity inversion line, enhancing the flux rope with a twist that is uniform along the rope but varies with radius. It also creates an arcade of rising flare loops and separating chromospheric ribbons.
Our simple model shows how the mean twist in the flux rope depends on the various geometrical properties of the pre-flare configuration, and we are also able to deduce the variation with flux of the internal twist inside the flux rope.

The new aspects are: a deduction of the amount of twist in the erupting flux rope from the initial geometry and the nature of the reconnection; the suggestion that the initial zipper phase of reconnection during the establishment of the flare loops and flare ribbons can build up strong core twist in the erupting flux rope; and a new relation between the resulting core twist and the ratio of the fluxes in the ribbons and the new part of the flux rope.

Interplanetary flux ropes of uniform twist could be produced either from an initial sheared arcade (provided it becomes eruptively unstable) or from an arcade containing an initial flux rope of moderate twist (up to one or two turns, which is more likely to become unstable). Interplanetary ropes of high twist, on the other hand, could be produced if the initial flux rope has high twist (in excess of two turns) 
or an active zipper phase creates many new turns around the initial flux rope.

In future, to help predict radial twist profiles in magnetic clouds, it would be useful to measure the preflare geometry and dimensions of the flare region, the twist in an initial flux rope before eruption, and the magnetic fluxes in the initial ribbons and the new initial flux rope as well as the total flux mapped out by the ribbons in their transverse motion.

Other important aspects to study include: determining whether the assumption of magnetic helicity equipartition is a good one or needs to be modified; comparing the model with observations and computational experiments, 
and making it more realistic; including the extra constraints from energy considerations; 
deducing the twist in an initial flux rope from observations; and determining the flux of the zipper flux rope by comparison with that of the overlying arcade and its effect on the erupting flux rope.  

\appendix
\section{Twist Produced by the First Zippette}

The twist produced by the first zippette from two initially untwisted flux tubes is, from Equation (\ref{eq:dPhi_g1}),
\begin{equation}
  \tan(\Phi_1) ~=~ -{2\bm\bar{L}^2\bm\bar{s}\over 9(\bm\bar{s}^2+1)^2 - (\bm\bar{s}^2-1)\bm\bar{L}^2},
  	\label{app1}
\end{equation}
where, when there is no shear initially ({\textit {i.e.}}, when ${\bm\bar s}=0$), we know that $\Phi_1=\pi$.

After writing $\bm\bar{L}^2=18a$ and $\tan(\Phi_1)=\bm\bar{f}=4a f$, the function $f({\bm\bar s},a)$ becomes
\begin{equation}
  f ({\bm\bar s},a)~\equiv~-\frac{{\bm\bar s}}{g({\bm\bar s},a)}~=~ -\frac{{\bm\bar s}}{{\bm\bar s}^4 +2(1-a){\bm\bar s}^2+1+2a}.
  	\label{app2}
\end{equation}

First of all, note that $f=0$ (and so $\Phi_1=\pi$) at ${\bm\bar s}=0$ and  as ${\bm\bar s}$ tends to $\pm \infty$. Also,  $f$ is an odd function of ${\bm\bar s}$, so that $\Phi_1-\pi$ is an antisymmetric function of ${\bm\bar s}$.
For values of $a$ for which $g({\bm\bar s},a)>0$ (namely, $0<a<4$, as we shall prove below),  $f<0$ when ${\bm\bar s}>0$ so that $\pi/2<\Phi_1<\pi$, whereas $f>0$ when ${\bm\bar s}<0$ so that $\pi<\Phi_1<3\pi/2$. On the other hand, if $g({\bm\bar s},a)$ dips below $0$ (which occurs when $a>4$), then there is a range of ${\bm\bar s}$ for which $0<\Phi_1<\pi/2$ for ${\bm\bar s}>0$ and $3\pi/2<\Phi_1<2\pi$ for ${\bm\bar s}<0$.

\begin{figure}[h]
{\centering
 \includegraphics[width=12cm]{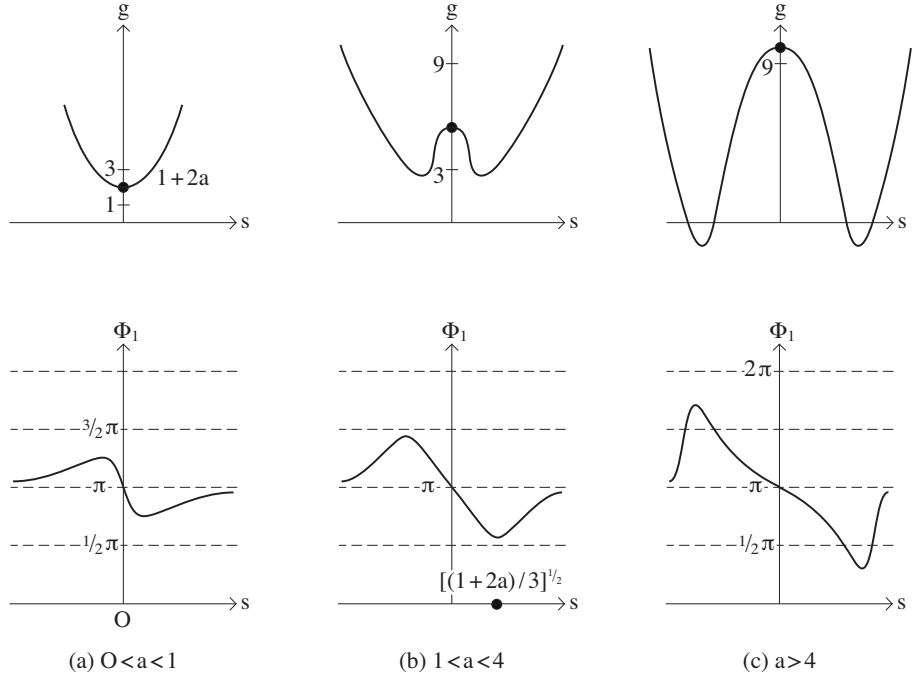}
\caption{The forms of $g({\bm\bar s},a)$ and the flux rope twist $\Phi_1({\bm\bar s},a)$ after one zippette as functions of ${\bm\bar s}$ for: (a) $0<a<1$ (\textit{i.e.}, $0<\bm\bar{L}<3\sqrt 2$; (b) $1<a<4$ (\textit{i.e.}, $3\sqrt 2<\bm\bar{L}<6\sqrt 2$); and  (c) $a>4$ (\textit{i.e.}, $\bm\bar{L}>6\sqrt 2$), where $\bm\bar{L}^2 =18a$, ${\bm\bar s}=s/w_z$ is the ratio of the arcade shear  to its width  in Figure \ref{fig6}, and ${\bm\bar L}=L/w_z$ is the ratio of the arcade length to its width.}
\label{fig14}}
\end{figure}

In order to establish these facts and sketch the curves, consider first $g({\bm\bar s},a)={\bm\bar s}^4 +2(1-a){\bm\bar s}^2+1+2a$, which is an even function of ${\bm\bar s}$, so we focus on its behaviour when ${\bm\bar s}\geq0$. At ${\bm\bar s}=0$, $g=1+2a$, which is always positive (since $a>0$) and increases with $a$, as indicated by the variation in the position of the large dot in the top row of Figure \ref{fig14}. Also, $g({\bm\bar s},a)$  is positive when $a<4$ but vanishes when $a>4$ at two positive values and two negative values given by
\begin{equation}
{\bm\bar s}^2=a-1\pm\sqrt{a(a-4)}.
\label{app3}
\end{equation}
The two positive values coincide at ${\bm\bar s}=\sqrt3$ when $a=4$.

Turning points of $g({\bm\bar s},a)$ as a function of ${\bm\bar s}$ for fixed $a$ are given by
\begin{equation}
\frac{\partial g}{\partial {\bm\bar s}}=4{\bm\bar s}({\bm\bar s}^2+1-a).
\label{app4}
\end{equation}
Thus, when $a<1$ there is only one turning point (a minimum) at ${\bm\bar s}=0$. However, when $a>1$ two more turning points appear at
\begin{equation}
{\bm\bar s}=\pm\sqrt{a-1},
\label{app5}
\end{equation}
which increases in magnitude with $a$.
The value of $g({\bm\bar s},a)$ at these two new turning points is
\begin{equation}
g_{\rm min}=a(4-a),
\label{app6}
\end{equation}
which represent minima, since there is only one turning point in the range ${\bm\bar s}>0$, $g(0,a)>0$ and $g\rightarrow +\infty$ as ${\bm\bar s}\rightarrow +\infty$. Thus, when $a>1$, the turning point at the origin becomes a maximum. The minimum value $g_{\rm min}$ is positive when $1<a<4$ but is negative when $a>4$, namely, when the four extra zeros of $g({\bm\bar s},a)$ appear.  Putting together this information, we arrive at the qualitative behaviour of $g({\bm\bar s},a)$ for different ranges of $a$ shown in the top row of Figure \ref{fig14}.

Next, consider $f({\bm\bar s},a)=\tan \Phi_1/(4a)$ and the corresponding form for the twist $\Phi_1$. First of all, note that $f$ vanishes only at ${\bm\bar s}=0$ and as ${\bm\bar s}\rightarrow \pm \infty$ and its gradient $\partial f/\partial {\bm\bar s}$ at the origin is $-1/(1+2a)$, which is always negative and decreases in magnitude as $a$ increases.

For $0<a<4$ (\textit{i.e.}, ${\bm\bar L}<6\sqrt 2$), $g>0$ and so $f$ is finite. It is negative when ${\bm\bar s}$ is positive and positive when ${\bm\bar s}$ is negative, so that $\pi/2<\Phi_1<3\pi/2$ (see the bottom row of Figure \ref{fig14}a and b).

Furthermore, $f({\bm\bar s},a)$ as a function of ${\bm\bar s}$ possesses a turning point where
\begin{equation}
\frac{\partial f}{\partial {\bm\bar s}}=-\frac{({\bm\bar s}^2+1)(-3{\bm\bar s}^2+1+2a)}{({\bm\bar s}^4+2(1-a){\bm\bar s}^2+1+2a)^2}
\label{app6}
\end{equation}
vanishes, namely, at 
\begin{equation}
{\bm\bar s}=\pm\sqrt{\frac{1+2a}{3}}.
\label{app7}
\end{equation}
 The value ($f_{\rm min}$) at ${\bm\bar s}=+\sqrt{(1+2a)/3}$ is 
\begin{equation}
f_{\rm min}=\frac{3\sqrt3}{4\sqrt{1+2a}(a-4)},
\label{app8}
\end{equation}
which is positive when $a>4$ and negative when $0<a<4$.

When $a>4$ (\textit{i.e.}, ${\bm\bar L}>6\sqrt 2$), in the region where ${\bm\bar s}>0$ there are two locations where $f$ becomes infinite, and between them $f$ becomes positive and possesses a minimum. These correspond to two locations where $\Phi_1$ passes through $\pi/2$, between which $\Phi_1$ possesses a minimum larger than 0. In the region ${\bm\bar s}<0$ there are also two locations where $f$ becomes infinite, between which $f$ is negative and possesses a maximum. These correspond to two locations where $\Phi_1$ passes through $3\pi/2$, between which $\Phi_1$ possesses a maximum smaller than $2\pi$.  These features are indicated on the bottom row of Figure \ref{fig14}c.

\acknowledgements
We are grateful to Mitch Berger, Pascal D{\' e}moulin, Miho Janvier, Clare Parnell and Jiong Qiu for helpful comments and suggestions and to the UK STFC for financial support.\\

\setlength{\parindent}{0cm}
{\bf Disclosure of Potential Conflicts of Interest}
 The authors declare that they have no conflicts of interest.


\clearpage

\end{article}
\end{document}